\documentclass[prd,twocolumn,aps,psfig,showpacs,nofootinbib,nobibnotes,superscriptaddress,preprintnumbers]{revtex4}

\usepackage{graphicx,natbib}
\usepackage[figuresright]{rotating}
\usepackage{amssymb}
\usepackage{amsmath}
\usepackage{bm}
\usepackage[usenames, dvipsnames]{color}
\usepackage{ulem}

\newcommand{\App}[1]{Appendix~\ref{#1}}
\newcommand{\Sec}[1]{Sec.~\ref{#1}}
\newcommand{\Eq}[1]{Eq.~(\ref{#1})}
\newcommand{\Eqs}[2]{Eqs.~(\ref{#1}) and (\ref{#2})}

\newcommand{\Eqss}[2]{Eqs~(\ref{#1})--(\ref{#2})}
\newcommand{\bra}[1]{\langle #1\rangle}
\newcommand{\nab}{\mbox{\boldmath $\nabla$} {}}
\newcommand{\DD}{{\rm{D}}}
\newcommand{\be}{\begin{equation}}
\newcommand{\ee}{\end{equation}}

\newcommand{\bea}{\begin{eqnarray}}
\newcommand{\eea}{\end{eqnarray}}
\newcommand{\bean}{\begin{eqnarray*}}
\newcommand{\eean}{\end{eqnarray*}}

\def\Rey{{\rm Re}}
\def\urms{u_{\rm rms}}

\def\Brms{B_{\rm rms}}
\def\cs{c_{\rm s}}
\def\betaM{\beta_{\rm M}}
\def\epsMz{\epsilon_{\rm M\star}}
\def\epsM{\epsilon_{\rm M}}

\def\sigmaM{\sigma_{\rm M}}
\def\sigmaMz{\sigma_{\rm M\star}}
\def\sigmaK{\sigma_{\rm K}}
\def\xiM{\xi_{\rm M}}
\def\xiMz{\xi_{\rm M\star}}
\def\xiK{\xi_{\rm K}}
\def\xiz{\xi_\star}
\def\pM{p_{\rm M}}

\def\PP{\bm{P}}
\def\PPM{\bm{P}_{\rm M}}
\def\PPK{\bm{P}_{\rm K}}
\newcommand{\SSSS}{\mbox{\boldmath ${\sf S}$} {}}
\newcommand{\AAA}{\bm{A}}
\newcommand{\BB}{\bm{B}}
\newcommand{\JJ}{\bm{J}}
\newcommand{\uu}{\bm{u}}
\newcommand{\rr}{\bm{r}}
\newcommand{\xx}{\bm{x}}
\newcommand{\kk}{\bm{k}}
\newcommand{\oo}{\bm{\omega}}
\def\EEK{{\cal E}_{\rm K}}
\def\EEM{{\cal E}_{\rm M}}

\def\HHM{{\cal H}_{\rm M}}
\def\HHC{{\cal H}_{\rm C}}
\def\EK{E_{\rm K}}
\def\EM{E_{\rm M}}

\def\HM{H_{\rm M}}
\def\HC{H_{\rm C}}
\def\la{\mathrel{\mathchoice {\vcenter{\offinterlineskip\halign{\hfil
$\displaystyle##$\hfil\cr<\cr\sim\cr}}}
{\vcenter{\offinterlineskip\halign{\hfil$\textstyle##$\hfil\cr<\cr\sim\cr}}}
{\vcenter{\offinterlineskip\halign{\hfil$\scriptstyle##$\hfil\cr<\cr\sim\cr}}}
{\vcenter{\offinterlineskip\halign{\hfil$\scriptscriptstyle##$\hfil\cr<\cr\sim\cr}}}}}
\def\ga{\mathrel{\mathchoice {\vcenter{\offinterlineskip\halign{\hfil
$\displaystyle##$\hfil\cr>\cr\sim\cr}}}
{\vcenter{\offinterlineskip\halign{\hfil$\textstyle##$\hfil\cr>\cr\sim\cr}}}
{\vcenter{\offinterlineskip\halign{\hfil$\scriptstyle##$\hfil\cr>\cr\sim\cr}}}
{\vcenter{\offinterlineskip\halign{\hfil$\scriptscriptstyle##$\hfil\cr>\cr\sim\cr}}}}}

\newcommand{\Fig}[1]{Fig.~\ref{#1}}
\newcommand{\FFig}[1]{Figure~\ref{#1}}

\newcommand{\Figs}[2]{Figs.~\ref{#1} and \ref{#2}}
\newcommand{\Figss}[2]{Figs.~\ref{#1}--\ref{#2}}
\newcommand{\Tab}[1]{Table~\ref{#1}}

\newcommand{\ypre}[3]{, Phys.\ Rev.\ E {\bf #2}, #3 (#1).}
\newcommand{\yprl}[3]{, Phys.\ Rev.\ Lett.\ {\bf #2}, #3 (#1).}
\newcommand{\yjfm}[3]{, J.\ Fluid Mech.\ {\bf #2}, #3 (#1).}

\newcommand{\yjour}[4]{ #1, {#2}, {#3}, #4}

\newcommand{\kpc}{\,{\rm kpc}}
\newcommand{\Mpc}{\,{\rm Mpc}}
\newcommand{\G}{\,{\rm G}}
\newcommand{\nG}{\,{\rm nG}}
\newcommand{\EQ}{\begin{equation}}
\newcommand{\EN}{\end{equation}}
\newcommand{\ba}{\begin{eqnarray}}
\newcommand{\ea}{\end{eqnarray}}

\def\ii{{\rm i}}
\def\const{{\rm const}}
\def\tauA{t_{\rm A}}
\def\tauAz{t_{\rm A\star}}
\def\vvA{\bm{v}_{\rm A}}
\def\vA{v_{\rm A}}
\def\vAz{v_{\rm A0}}
\def\xiM{\xi_{\rm M}}
\def\HHM{{\cal H}_{\rm M}}
\def\kB{k_{\rm B}}
\graphicspath{{./fig/}{./png/}}

\begin{document}
\title{
Evolution of hydromagnetic turbulence from the electroweak phase transition
}

\preprint{NORDITA-2017-116}

\author{Axel~Brandenburg}
\affiliation{Laboratory for Atmospheric and Space Physics, University of Colorado, Boulder, CO 80303, USA}
\affiliation{JILA and Department of Astrophysical and Planetary Sciences, University of Colorado, Boulder, CO 80303, USA}
\affiliation{Nordita, KTH Royal Institute of Technology and Stockholm University,
Roslagstullsbacken 23, 10691 Stockholm, Sweden}
\affiliation{Department of Astronomy, AlbaNova University Center,
Stockholm University, 10691 Stockholm, Sweden}

\author{Tina~Kahniashvili\footnote{Corresponding author; the authors are listed alphabetically.}
}
\email{tinatin@andrew.cmu.edu}
\affiliation{McWilliams Center for
Cosmology and Department of Physics, Carnegie Mellon University,
5000 Forbes Ave, Pittsburgh, PA 15213, USA}
\affiliation{Department of Physics, Laurentian University, Ramsey
Lake Road, Sudbury, ON P3E 2C,Canada}
\affiliation{Abastumani Astrophysical Observatory, Ilia State University,
3-5 Cholokashvili St., 0194 Tbilisi, Georgia}

\author{Sayan~Mandal}
\affiliation{McWilliams Center for
Cosmology and Department of Physics, Carnegie Mellon University,
5000 Forbes Ave, Pittsburgh, PA 15213, USA}

\author{Alberto~Roper~Pol}
\affiliation{Laboratory for Atmospheric and Space Physics, University of Colorado, Boulder, CO 80303, USA}
\affiliation{Department of Aerospace Engineering Sciences, University of Colorado, Boulder, CO 80303, USA}

\author{Alexander~G.~Tevzadze}
\affiliation{Faculty of Exact and Natural Sciences, Javakhishvili Tbilisi
State University, 3 Chavchavadze Ave., Tbilisi, 0179, Georgia}
\affiliation{Abastumani Astrophysical Observatory, Ilia State University, 3-5
Cholokashvili St., 0194 Tbilisi, Georgia}

\author{Tanmay~Vachaspati}
\affiliation{Physics Department, Arizona State University,
Tempe, AZ 85287, USA}

\date{\today
}
\begin{abstract}
We present new simulations of decaying hydromagnetic turbulence for a
relativistic equation of state relevant to the early universe.
We compare helical and nonhelical cases either with kinetically or
magnetically dominated initial fields.
Both kinetic and magnetic initial helicities lead to maximally helical
magnetic fields after some time, but with different temporal decay laws.
Both are relevant to the early universe, although no mechanisms have yet
been identified that produce magnetic helicity with strengths comparable
to the big bang nucleosynthesis limit at scales comparable to the Hubble
horizon at the electroweak phase transition.
Nonhelical magnetically dominated fields could still produce
picoGauss magnetic fields under most optimistic conditions.
Only helical magnetic fields can potentially have nanoGauss strengths
at scales up to $30\kpc$ today.
\end{abstract}

\maketitle

\section{Introduction}
\label{sec:introduction}
A host of astrophysical observations indicate the presence of coherent
magnetic fields with strengths at the
microGauss ($\mu$G) level from
the scale of galaxies to clusters of galaxies \cite{Widrow:2002ud}.
It is thought that such fields may have originated from cosmological or
astrophysical seed fields which were subsequently amplified during structure
formation, via processes like adiabatic compression and MHD turbulence
instabilities \cite{Kulsrud:2007an,Kandus:2010nw,Durrer:2013pga}.
The statistical properties of the resulting magnetic field, viz. the
amplitude, spectral shape, and the correlation length, depend strongly on the
initial conditions, i.e., on the particular generation mechanisms.

Primordial magnetic fields can be generated through \textit{causal}
processes which include all astrophysical scenarios as well as primordial
magnetogenesis occurring after inflation.
In all those cases, the correlation length is bounded and limited by
the {\it causal horizon} which is associated with the Hubble horizon
scale at the time of magnetic field production \cite{Hogan:1983zz}.
If one accounts for the turbulent magnetic evolution during the expansion
of the universe, the correlation length may reach galactic length scales
today \cite{Kahniashvili:2012uj}.
In contrast, Refs.~\cite{Wagstaff:2014fla,Reppin:2017uud} assumed that the turbulent evolution is
less effective in increasing the magnetic correlation length and obtained a faster decay of
the magnetic field energy.

The evolution of the magnetic field as the universe expands, as well as other observable signatures, depend strongly on the magnetic helicity
of the initial seed field \cite{Banerjee:2004}.
A number of astrophysical objects, ranging from stars
\cite{Brandenburg:2004jv} to jets from active galactic nuclei
have detectable magnetic helicity \cite{Ensslin:2002gn}.
Usually, the magnetic helicity is initially much less than the maximum
possible value, which is given by the product of magnetic energy and
the magnetic correlation length.
However, the fractional helicity increases due to MHD turbulence.
This leads to a maximally helical configuration of the observed fields~\cite{Tevzadze:2012kk}.

If primordial magnetic helicity is detected, it will indicate a violation of parity (mirror symmetry) violation in the early
universe, and may point towards a resolution to the matter-antimatter asymmetry
problem~\cite{Long:2013tha,Fujita:2016igl,Kamada:2016eeb,Kamada:2016cnb}.
To generate causal helical magnetic fields in the early universe, one requires fundamental
parity violation that affects the outcome of cosmological phase
(electroweak or QCD) transitions \cite{Cornwall:1997ms,Giovannini:1997eg,Vachaspati:1991nm,
Joyce:1997uy,Field:1998hi,Vachaspati:2001nb,Chu:2011tx,Tashiro:2012mf,Copi:2008he,
Sigl:2002kt,Subramanian:2004uf,DiazGil:2007dy,Stevens:2007ep,Forbes:2000gr,
Boyarsky:2011uy,Ng:2010mt}.

Assuming that the (comoving) mean energy density of the magnetic
field $\EEM$ ($\equiv\bra{\BB^2}/2$, where
$\BB$ is the magnetic field strength) depends only on the present day
temperature $T_0$ and fundamental constants such as the Boltzmann constant
$\kB$, the reduced Planck constant $\hbar$, and the speed of light $c$,
one finds, on dimensional grounds,
\EQ
\bra{\BB^2}/2\la\epsilon_1(\kB T_0)^4/(\hbar c)^3,
\label{argument2}
\EN
where $\epsilon_1$ is a dimensionless number. For $\epsilon_1=1$, this results
in a root mean square (rms) field strength of $3\times10^{-6}\G$.
(To get the field in gauss, one has to multiply the Lorentz-Heaviside
value by $\sqrt{4\pi}$.)
A certain fraction of this magnetic field strength is also what is known
as the big bang nucleosynthesis bound (BBN), which implies that the total
energy density budget, in addition to radiation and other relativistic
components, should not exceed 10\% of the radiation energy density; see
Sec.~\ref{subsec:initialconditions}. Eq.~(\ref{argument2}) implies that
the mean comoving magnetic energy density is determined by today's
temperature $T_0$. On the other hand, today's temperature is set by
the photon (radiation) energy density, and the dimensionless quantity
$\epsilon_1$ is a ratio between the mean comoving magnetic energy and
today's radiation energy densities.

The conserved magnetic helicity per unit volume, i.e., the mean magnetic
helicity density\footnote{In the following we talk about magnetic helicity
and omit the specification to mean helicity density for simplicity.}
is roughly given by $\bra{\BB^2}\,\xiM$, where
$\xiM$ is the magnetic correlation length.
As above, assuming that this product depends only on $T_0$ and
the fundamental constants $\kB$, $\hbar$, and $c$, one finds
\cite{Brandenburg:2017rcb}
\EQ
\bra{\BB^2}\,\xiM\la\epsilon_2(\kB T_0)^3/(\hbar c)^2,
\label{argument}
\EN
where $\epsilon_2$ is a dimensionless number.
This results in a field strength of $5\times10^{-19}\G$ for $\xiM=1\Mpc$
and $\epsilon_2=1$.
(For $\xiM=10\kpc$, which is more suitable for magnetic fields produced
during the electroweak phase transition \cite{Kahniashvili:2012uj},
the corresponding field strength would be $5\times10^{-18}\G$.)

Larger values of $\bra{\BB^2}\,\xiM$ are possible if the underlying
physics involves another fundamental constant,
for example Newton's constant $G$.
In that case, again just on dimensional grounds, one can write
\EQ
\bra{\BB^2}\,\xiM\la\epsilon_3 \,(a_\star/a_0)^3\,
G^{-3/2}\hbar^{-1/2}c^{11/2},
\label{argument3}
\EN
where $a_\star/a_0=8\times 10^{-16}$ is the ratio of the scale factor
at the time of magnetic field generation (the electroweak phase transition)
to that at the present time.
This corresponds to a field strength of $4\times10^6\G$ for $\xiM=1\Mpc$
and $\epsilon_3=1$.
Alternatively, of course, geometric means between \Eqs{argument}{argument3}
are conceivable.
Of particular interest would be a $2:1$ mixing ratio,
\EQ
\bra{\BB^2}\,\xiM\la\epsilon_2^{2/3} \epsilon_3^{1/3} (a_\star/a_0)\, (\kB T_0)^2
\,G^{-1/2}\hbar^{-3/2}c^{1/2}, \label{argument23}
\EN
i.e., $10^{-20}\epsilon_2^{2/3}\epsilon_3^{1/3}\G^2\Mpc$,
or $10^{-10}\G$ for $\xiM=1\Mpc$ and $\epsilon_2=\epsilon_3=1$.
This mixing ratio corresponds to the magnetic field being at
the BBN limit and $\xiM$ being comparable to the Hubble scale.

The considerations above do not allow us to predict the
maximum available magnetic helicity unless some physical
mechanism is unidentified.
In the case of the chiral magnetic effect \cite{Joyce:1997uy,Boyarsky:2011uy},
for example, Newton's constant does not enter, and so \Eq{argument}
does impose a rather stringent constraint.
However, if stronger magnetic helicities are to be produced by
some as yet unknown mechanism, this should allow us to identify a
nonvanishing mixing ratio between \Eqs{argument}{argument3}.
The ratio 2:1 is physically appealing, but by no means the only possible
choice.
Note, however, that the 2:1 ratio is also being reflected in the
magnetogenesis scenario with a strong charge-parity (CP) violation.
One such option is presented by the scenario of Ref.~\cite{Forbes:2000gr},
in which maximal helicity is produced through Chern-Simons CP violation
leading to magnetic fields correlated on $100\kpc$ scales.

In this paper we focus on magnetogenesis mechanisms during the electroweak phase transition,
as proposed in Refs.~\cite{Vachaspati:1991nm,Cornwall:1997ms,Vachaspati:2001nb,
Copi:2008he,Ng:2010mt},
assuming that the phase transition is strongly first order.
Our main goal is to study the dynamical evolution of the generated magnetic fields during
the expansion of the universe and estimate if they can
serve as the initial seed for the observed magnetic fields in galaxies
and clusters.
We will determine the evolution of the magnetic
field from the electroweak epoch until the epoch of recombination.
We can evolve the magnetic field from recombination to the present epoch by using the fact
that the primordial plasma is neutral after recombination
and the free MHD decay stops, so the
comoving amplitude, spectral shape, and helicity of the magnetic field stay unchanged until large scale structure formation and reionization. In the following, we neglect further nonlinear evolution
of the magnetic field during large-scale structure formation and re-ionization.

Since the first order phase transition proceeds via bubble nucleation and subsequent
collision of these bubbles \cite{Nicolis:2003tg}, there is stirring of the plasma at high Reynolds number, and a consequent generation of turbulent motion.
This occurs in addition to the magnetic fields that are produced.
Correspondingly, the turbulent motions can be (i) magnetically dominant, (ii) hydrodynamically dominant (i.e. magnetically subdominant), or (iii) have an equipartition between magnetic
field and velocity field energies. We address all these cases separately.
Most of the earlier investigations have employed magnetically dominated turbulence
\cite{CHB01,Banerjee:2004}.
Recently, for the first time, we have considered a
magnetically subdominant case, but with initial kinetic helicity
\cite{Brandenburg:2017rnt}.\footnote{The generation of kinetic helicity during parity
or chirality violating electroweak phase transitions can be expected
since the interaction strengths of the left- and right-handed particles
are different.}

In Sec.~\ref{sec:magnetogenesis} we briefly review the electroweak phase transition
magnetogenesis, and we determine the initial conditions for further
evolution of the magnetic field.
We study the evolution of MHD turbulence under the initial conditions
presented in Sec.~\ref{sec:magnetogenesis} using direct numerical simulations (DNS) in
Sec.~\ref{sec:MFEvolution}.
We discuss our results and conclude in Sec.~\ref{sec:results}.
From now on, we use natural ($\hbar = k_B= c=1$) Lorentz-Heaviside units.
So there are no factors of $4\pi$ in Maxwell equations and magnetic
energy density is $\BB^2/2$.
Unless specified, $t$ denotes the conformal time, $dt = d\tau/a(\tau)$
(with $\tau$ the physical time, and $a=a(\tau)$ the scale factor).
We normalize the scale factor to be unity today, i.e.\ $a_0=a(\tau=\tau_0)=1$.

The expansion of the universe can be eliminated from the relativistic MHD equations
through use of suitably re-scaled (comoving) quantities \cite{Brandenburg:1996fc}.
For example, we use the comoving value for the magnetic field, i.e.\ ${\bm B} \rightarrow a^2{\bm B}$
that also reflects magnetic flux conservation for a {\it frozen-in} magnetic
field in the expanding universe.
To avoid confusion ${\tilde {\bm B}}$ will denote the physical magnetic field.

\section{Electroweak Phase Transition Magnetogenesis}
\label{sec:magnetogenesis}

We investigate the scenario where a cosmological magnetic field is
generated during baryogenesis at the electroweak phase transition
at conformal time $t=t_\star$ (that corresponds to the temperature $T_\star$). The phase transition is assumed
to be strongly first order, and the magnetic field is produced by anomalous
baryon number violation as described in
Refs.~\cite{Vachaspati:2001nb,Ng:2010mt,Copi:2008he,Chu:2011tx,
Vachaspati:1991nm,Vachaspati:1994xc}.
The magnetic field immediately after production is assumed to be statistically homogeneous and isotropic, Gaussian-distributed vector field,
and is described in terms of
the equal time correlation function \cite{my75},
\begin{equation}
\langle B_i^*({\bm k}, t) B_j({\bm k}', t) \rangle = (2\pi)^3 \delta^{(3)}({\bm k}-{\bm k}') F_{ij} ({\bm k}, t),
 \label{equaltimespectrum}
\end{equation}
where ${\bm B}({\bm k},t)$ is the Fourier transform\footnote{We use the following
convention for the forward and inverse Fourier transform of an arbitrary vector field ${\bm A}(\bm x)$
\begin{equation}\nonumber
\begin{aligned}
A_i({\bm k}) &= \int d^3x \, A_i({\bm x})\,e^{+i{\bm k}\cdot {\bm x}},\\
A_i({\bm x}) &= \int \frac{d^3k}{(2\pi)^3} A_i({\bm k})\,e^{-i{\bm k}\cdot {\bm x}}
\end{aligned}
\end{equation}
}
of ${\bm B}({\bm x},t)$. The correlation function $F_{ij}({\bm k}, t)$
has nonhelical (symmetric) and helical (antisymmetric) components,
\begin{eqnarray}
\frac{F_{ij}({\bm k}, t)}{(2\pi)^3}
= P_{ij} ({\bm \hat k}) \frac{\EM(k,t)}{4\pi k^2}
+ i \epsilon_{ijl} k_l \frac{\HM (k,t)}{8\pi k^2},
\label{equaltimeF-function}
\end{eqnarray}
where $P_{ij}({\bm \hat k}) \equiv \delta_{ij}-{\hat k}_i{\hat k}_j$ is the projection operator that projects
any vector in the direction orthogonal to ${\bm {\hat k}}$
and ensures the solenoidal nature of the magnetic field.

Note that the form of the correlation function in
Eq.~(\ref{equaltimespectrum}) assumes statistical isotropy --
rotational symmetry is preserved, while mirror (parity) symmetry is broken by the
helical component.
Assuming that the real space two-point correlation function
$\langle {\bm B}({\bm x}) {\bm B} ({\bm x+r}) \rangle$ vanishes for $|{r}| \rightarrow \infty$, the form of the correlator $F_{ij}({k},t)$
in \Eq{equaltimeF-function} is strictly valid only if
the spectrum $\EM(k,t)$ falls off faster than $k^2$ as $k \to 0$
and fixed time $t$
\cite{my75}.\footnote{The {\it causal} magnetogenesis mechanisms considered here do not
include magnetic fields generated during cosmological inflation in which a scale invariant
spectrum with
$\EM(k) \propto k^{-1}$ is produced.
A scale-invariant spectrum has an {\it unlimited} correlation length scale and cannot
be generated by causal processes during cosmological phase transitions.
Following Ref.~\cite{my75}, the requirement that the correlation function in
Eq.~(\ref{equaltimespectrum}) be analytic
for $k \rightarrow 0$ leads to $\EM(k)
\propto k^4$ (so called
Batchelor spectrum). A similar shape has been discussed in Ref.~\cite{Durrer:2003ja} in which the authors
argued that the magnetic field should have strictly vanishing spatial correlation on length scales larger than
the cosmological horizon scale and then should fall off faster than $k^4$ (instead of $k^2$ - white noise) to
be divergence free.}

\subsection{Modeling Primordial Magnetic Field}
\label{subsec:initialconditions}

Motivated by electroweak baryogenesis, extensions of the standard model
in which the electroweak phase transition is strongly first order have been considered
(recently in~\cite{Huang:2016cjm}).
The models include the standard model with an extra
singlet~\cite{Espinosa:2011ax}, the two-Higgs doublet model~\cite{Kozaczuk:2014kva},
and the Next-to-Minimal Supersymmetric Standard Model (NMSSM)~\cite{Kozaczuk:2014kva}.
For our work we will assume that there is a strong first order phase transition
at the electroweak epoch \cite{Espinosa:2010hh}. The phase transition then proceeds by bubble nucleation and growth, and since it is a strong first
order transition, the typical bubble size at percolation can be large, perhaps even of the order of
$\tau_\star$.
During the phase transition,
there are baryon number violating particle interactions in the
medium that also generate helical magnetic fields as a
by-product \cite{Vachaspati:2001nb,Copi:2008he,Chu:2011tx}. Far outside the bubbles, where
electroweak symmetry is unbroken, we expect the magnetic fields to be in thermal equilibrium. Inside
the bubbles, the electroweak symmetry is broken, the weak gauge fields are massive, and
baryon number violation is suppressed. Then there is no magnetic field production within the bubbles.
However, any magnetic field that is generated just outside the bubble walls gets trapped once the bubble
expands further and this magnetic field can survive. Once the phase transition is over, space is filled
with helical magnetic fields that were generated by baryon number violation occurring near the bubble
walls.

Baryon number violating processes will sometimes produce baryons and sometimes anti-baryons.
CP violation in the model will yield a slight excess of baryons.
In terms of magnetic fields, this means that both left- and right-handed
magnetic fields will be produced but there will be an excess of
left-handed helicity.

A strong first order electroweak phase transition is also likely to produce turbulence in the cosmological
medium \cite{Nicolis:2003tg}.
Particles of the cosmological medium are massless outside the bubbles and massive within.
Thus the bubble wall interacts with the particles and pushes the medium in front of it in what is
described as a snowplow effect.
The typical turbulence eddy turnover velocity is given by,
\begin{equation}
u_{T} = \sqrt{\frac{\tilde\kappa\,\tilde\alpha}{\frac{4}{3}+\tilde\kappa\,\tilde\alpha}},
\end{equation}
where $\tilde\alpha$ denotes the ratio of the false vacuum energy density (latent heat) and the plasma thermal
energy density, and characterizes the strength of the phase transition; $\tilde\kappa$ is an efficiency
parameter that is determined by $\tilde\alpha$ and has to be computed numerically
\cite{Kamionkowski:1993fg},
\begin{equation}
\tilde\kappa(\tilde\alpha) \simeq \frac{1}{1+0.715\,\tilde\alpha}\left [\, 0.715\,\tilde\alpha +\frac{4}{27}\sqrt{\frac{3\tilde\alpha}{2}}\, \right ].
\end{equation}
A strong phase transition is described by $\tilde\alpha \gtrsim 1$ and a weak phase transition has
$\tilde\alpha \ll 1$.

Another important parameter that characterizes the forcing stage
of turbulence is the duration of the phase transition described
by a parameter $\tilde\beta$, which is the rate of time variation
of the nucleation rate itself computed at the phase transition time
$\tau_\star$, i.e., $\tilde\beta^{-1}$ gives a time scale during which
the whole universe is converted to the true vacuum phase (typically
$\tilde\beta \gg H_\star$) \cite{Kosowsky:2001xp}.

An outcome of the direct simulations of the magnetic field generation process is that the initial
magnetic field spectrum is peaked at a scale that corresponds to the size of the bubbles at percolation.
Hence, for a strong first order phase transition, the initial magnetic
field can be correlated on cosmological scales.
Let us denote this initial (physical) correlation
length by $l_\star$ and define the dimensionless parameter $\gamma_\star = l_\star H_\star$, which
we will take as a free parameter in the interval $10^{-4} < \gamma_\star < 0.1$.
It is of interest to evaluate the comoving value of the Hubble length scale at the electroweak phase transition.
We have already stated $H_\star^{-1} \approx 1~{\rm cm}$. Then the comoving value, denoted
$\lambda_{H_\star}$, is given by
\begin{equation}
\lambda_{H_\star} \equiv \frac{a_0}{a_\star} H_\star^{-1} = 5.8 \times 10^{-10}~{\rm Mpc}
\left(\frac{100\,{\rm GeV}}{T_\star}\right)
\left(\frac{100}{g_\star}\right)^{{1}/{6}},
\label{lambda-max}
\end{equation}
where the subscripts $\star$ and $0$ denote respectively the epoch of the magnetic field
generation and the present epoch; $g_\star$
is the number of relativistic degrees of freedom in the medium at the electroweak epoch,
and we have used the time-temperature relation
 \be
 \frac{a_\star}{a_0} \simeq 8 \times 10^{-16} \left(\frac{100~{\rm GeV}}{T_\star} \right)
 \left(\frac{100}{g_\star}\right)^{1/3}.
 \label{scalling-factor}
 \ee
The numerical value of $\lambda_{H\star} \approx 6 \times 10^{-4}~{\rm pc}$
is much smaller than the current horizon scale $\sim 1~{\rm Gpc}$, and
without significant growth, would not be an interesting scale for
astrophysics. However, it is known that turbulent MHD evolution of helical
magnetic fields allows for an inverse cascade that can lead to a significantly
larger coherence scale, even larger than $\sim 10~{\rm kpc}$
\cite{Copi:2008he,Kahniashvili:2012uj}.
This is also seen in the results of our simulations.

An important
quantity associated with the primordial magnetic field is its
total energy density at the moment of generation, $\rho_{\rm M \star}$.
Since the frozen-in
{(physical)} magnetic field amplitudes scales with
the expansion of the universe as ${\tilde B}\sim a^{-2}$,
the magnetic energy density scales like {\it radiation}
with the expansion of the universe if dissipation and/or amplification
processes are ignored. So the ratio of the magnetic and radiation
energy densities stays constant during the expansion of the universe.
BBN bounds the radiation-like energy density
during BBN, and {\it only} $\sim 10\%$ of the ordinary radiation
energy density can be additionally present in the universe in the
form of another relativistic component \cite{Grasso:2000wj}.
In particular, during radiation-dominated epoch, neglecting the presence of any additional relativistic components,
the Friedman first equation in the flat Friedmann-Lema\^itre-Robertson-Walker (FLRW) metric, reads
$
3H^2 = {8\pi G} \rho_{\rm R}
$
where $\rho_{\rm R}$ denotes the (physical) radiation energy density.
The expansion rate ($H$) can be limited by the rate of the nucleosynthesis
(that is bounded by the abundance of the light elements in the universe).
At the electroweak epoch, the radiation energy density is given by
$
\rho_{R}(t_\star) = {\pi} g_\star T_\star^4/30
$
with $g_\star$ are the degrees of freedom at the temperature $T_\star$ with ``$\star$'' subscript referring to the
moment of the magnetic field generation (i.e. the electroweak epoch).
Applying the BBN bound that ${\rho}_{\rm M}(t_\star)/ \rho_{R}(t_\star) \leq 0.1$ (with $\rho_{\rm M} = {\tilde B}^2/2$) and
assuming a frozen-in magnetic field (${\tilde B} \propto a^{-2}$), the {\it comoving} magnetic field strength can be
no larger than $ 8.4 \cdot 10^{-7} (100/g)^{1/6}{\rm G} \sim 1~\mu{\rm G}$,\footnote{
Here we use that the number of the relativistic degrees $g$ of freedom is unchanged from electroweak phase
transition till BBN.} which agrees well with the dimensional argument given in \Sec{sec:introduction}.
In our simulations we will take
\begin{equation}
b_\star \equiv \sqrt{\frac{\rho_{\rm M\star}}{0.1\rho_{R\star}}} \approx \frac{B_\star}{\mu{\rm G}} \lesssim 1
\label{bstar}
\end{equation}
to be a free parameter of the model.

We can define the Alfv\'en velocity associated with the magnetic field
$\vvA=\BB/\sqrt{w}$
where $w=\rho+p$ represents the enthalpy
of the fluid with density $\rho$ and pressure $p$. Then the normalized magnetic energy
density $\langle\vvA^2(t)\rangle/2$
allows a full analogy with the kinetic energy
density \cite{landau}, $\langle\uu^2(t)\rangle/2$ with $\uu(\xx,t)$ denoting
the velocity field, {where the angular brackets denote ensemble averaging.}

If the physical magnetic field scales as $a^{-2}$ with the expansion of the universe,
the Alfv\'en velocity
$\vvA(\xx,t)$ is time independent, and thus does not require
re-scaling to the comoving quantity.
At this point $\vvA(\xx,t)$ is fully determined by the initial value of the magnetic field,
i.e., $\vA(\xx,t) = \vAz(\xx) \equiv\vA(\xx,t=t_\star)$.

Owing to the presence of hydromagnetic turbulence,
the magnetic field evolution can be described by a
simple power law function, $B({\bm x}, t) = B_\star ({\bm x}) (t/t_\star)^{n_E/2}$ where $n_E$
determines the scaling of the magnetic field
energy density $\EM(k,t)$ decay.

The mean magnetic energy density
$\EEM(t)=\langle{\bm B}^2({\bm x},t)\rangle/2$ can be written in
terms of the magnetic spectral energy density $\EM(k,t)$ as
\begin{equation}
\EEM(t) = \int {dk}\,\EM(k,t),
\end{equation}
while the magnetic helicity\footnote{Note that $\HHM(t)$ differs
from the commonly used {\it current helicity}
$\HHC(t) = \langle {\bm B}\cdot {\rm {\bm curl}}({\bm B}) \rangle$.
We also define the current helicity spectral density $\HC(k,t) \equiv k^{2}\HM(k,t)$.},
defined as $\HHM = \langle\mathbf{A}\cdot\mathbf{B}\rangle$
with $\mathbf{B}=\nabla\times{\bm A}$, and
can be computed through the spectral helicity density as
\begin{equation}
\HHM(t) = \int dk\,\HM(k,t).
\end{equation}
The magnetic correlation length is defined as
\begin{equation}
\xiM(t) = \frac{\int dk \, k^{-1} \EM(k,t)}{\EEM(t)}~.
\label{correlation-length}
\end{equation}
Assuming that this integral is defined, the realizability condition
can be written as
\begin{equation}
2\EM(k,\tau) \geq k |\HM(k,\tau)|.
\end{equation}
This is a consequence of the Cauchy-Buyanovsky-Schwarz inequality and
implies that the magnetic energy cannot decay faster than the helicity
\cite{Candelaresi:2011pg}.
On integration, the realizability condition gives
\cite{Arnold1974,Moffat85}
\begin{equation}
2 \xiM(\tau) \EEM(\tau) \geq |\HHM(\tau)|
\label{realizability}
\end{equation}
and implies that the {\it maximal} helicity is
$2\int_0^\infty dk k^{-1} \EM(k)$.

Alternately, one can say that there is a lower bound on $\xiM$ given by,
\begin{equation}
\xiM^{\rm min}(t) \equiv \frac{ \HHM(t)}{2 \EEM(t)}
\label{xiMdefn}
\end{equation}
The realizability condition then implies
$\xiM^{\rm min} \leq \xiM$.
This allows us to define the fractional magnetic helicity as
\begin{equation}
\epsM(t)=\frac{ \xiM^{\rm min}(t) }{ \xiM(t) }
=\frac{ \HHM(t)}{2 \xiM(t) \EEM(t)} \leq 1.
\label{sigma}
\end{equation}
Its initial value is related to a parameter $\sigmaMz$ that will
be defined below and will serve as a free parameter; see below.

Another free parameter in our consideration is initial velocity field, $u_\star$. Applying the BBN bound
on the kinetic energy density ${\mathcal E}_K(t)$ that should be less than 10\% of the radiation energy
density (i.e., $\leq 0.1 \rho_{\rm R}$), we obtain that $u_\star \leq 0.4$ if decay and/or
amplification of the velocity field during turbulence
is neglected (the initial velocity field is assumed to be unchanged from the electroweak epoch until the BBN
epoch).

\section{Magnetic Field Evolution}
\label{sec:MFEvolution}

We follow the evolution of fields from
the epoch right after magnetogenesis up to the recombination epoch.
We are interested in the evolution of the magnetic energy density
$\EEM(t)$, which determines the rms value of the magnetic field,
$B_{\rm rms}(t)=\sqrt{2\EEM(t)}$,
and the correlation length $\xiM(t)$
For a partially helical magnetic field we
study re-distribution of helical structure at large scales, and estimate the time scale during which the field might
become fully helical. We also study the evolution of the velocity field.

\subsection{Direct Numerical Simulations}

We solve the equations for the logarithmic total energy density
$\ln\rho$, the velocity $\uu$,
and the magnetic vector potential $\bm{A}$ in the form \cite{Brandenburg:1996fc}
\begin{equation}
{\partial\ln\rho\over\partial t}=
-\frac{4}{3}\left(\nab\cdot\uu+\uu\cdot\nab\ln\rho\right)
+{1\over\rho}\left[\uu\cdot(\JJ\times\BB)+\eta\JJ^2\right]
\label{dlnrhodt}
\end{equation}
\vspace{-6mm}
\begin{eqnarray}
{\DD\uu\over\DD t}\!\!&=&\!\!
{\uu\over3}\left(\nab\cdot\uu+\uu\cdot\nab\ln\rho\right)
-{\uu\over\rho}\left[\uu\cdot(\JJ\times\BB)+\eta\JJ^2\right]
\nonumber
\\
&&-{1\over4}\nab\ln\rho
+{3\over4\rho}\JJ\times\BB+{2\over\rho}\nab\cdot\left(\rho\nu\SSSS\right)
\label{dudt}
\end{eqnarray}
\vspace{-6mm}
\begin{equation}
{\partial\BB\over\partial t}=\nabla\times(\uu\times\BB-\eta\JJ),
\label{mhd3}
\end{equation}
where ${\bm B}=\nabla \times {\bm A}$ and
$\DD/\DD t=\partial/\partial t +\bm{u}\cdot\bm{\nabla}$
is the advective derivative,
$\bm{f}_{\rm visc}=\nu\left(\nabla^2{\bm u}
+{\textstyle\frac{1}{3}}\bm{\nabla}\bm{\nabla}\cdot\bm{u}+\bm{G}\right)$ is
the viscous force in the compressible case with
$G_i=2{\sf S}_{ij}\nabla_j\ln\nu\rho$ as well as
${\sf S}_{ij}=\frac{1}{2}(u_{i,j}+u_{j,i})-\frac{1}{3}\delta_{ij}u_{k,k}$
being the
trace-free rate of strain tensor. The pressure is given by $p=\rho\cs^2$,
where $\cs=1/\sqrt{3}$ is the sound speed for an ultra-relativistic gas.
Furthermore, $\bm{J}=\bm{\nabla}\times\bm{B}$ is the current density.
In \App{ComparisonStandardMHD} we discuss the main difference to the
usual MHD equations for a non-relativistic isothermal gas.

In contrast to some of our previous studies
\cite{Kahniashvili:2010gp,Kahniashvili:2012uj,Tevzadze:2012kk} in
which the initial spectrum of magnetic field has been assumed to be a
$\delta$-function (the magnetic field energy density has been injected
at a given scale), in the present work we assume the initial spectral
distribution given by $\EM(k,t_\star)$ and $\HM(k,t_\star)$.
We also use different conditions for the velocity field, including the
magnetically subdominant case and equipartition with the magnetic field.
The magnetically dominant case has been studied previously, see
\cite{Brandenburg:2016odr} and references therein, but the magnetically
subdominant and equipartition cases have not been studied. In particular, the case of kinetically dominant and equipartition MHD
decay is presented for the first time in the recent publication \cite{Brandenburg:2017rnt} by
the authors and its application to cosmology is discussed below.

We allow $\nu$ and $\eta$ to be time-dependent; see
Ref.~\cite{Brandenburg:2017rnt} for details.
This is done to address the problem that $\nu$ and $\eta$
are very small in the early universe, but are also subject
to numerical limitations in that they cannot be too small, especially
at early times when the velocities are still large.
We take advantage of the fact that a self-similar evolution is possible
by allowing $\nu$ and $\eta$ to vary as
\begin{equation}
\nu(t)=\nu_\star\max(t,t_\star)^r,
\end{equation}
where $r=(1-\alpha)/(3+\alpha)$ \cite{Ole97}
depends on the initial power law slope $\alpha$,
and $t_\star$ is the minimal time after which these coefficients are allowed
to be time-dependent.
For $\alpha=2$ we have $r=-0.20$, while for $\alpha=4$ we have $r=-0.43$,
so $\nu(t)$ decreases with time in both cases.
We take different values of $\nu_\star$, depending on the value of $\alpha$.
In all cases with $\alpha=2$ we use $\nu_\star=10^{-6}$, while
in all cases with $\alpha=4$ we use $\nu_\star=10^{-5}$.
We adopt the same initial values for $\eta$, i.e., $\eta_\star=\nu_\star$.

For our numerical simulations we use the {\sc Pencil Code}
(\url{https://github.com/pencil-code})
which is a public MHD code that is particularly
well suited for simulating turbulence.
We consider a cubic domain of size $L^3$, so the smallest
wave number in the domain is $k_1=2\pi/L$.
The numerical resolution is $1152^3$ meshpoints in all the cases
presented below.

\subsection{Initial condition}
\label{sec:initialcondition}

In practice, we construct the initial condition for the magnetic vector potential
${\bm A}({\bm x})$ from a random three-dimensional vector field in real space
that is $\delta$-correlated. It has therefore a $k^2$ spectrum.
We transform this field into Fourier space and construct the magnetic field,
${\bm B}({\bm k}) = i{\bm k}\times {\bm A}({\bm k})$. We then
scale the magnetic field by functions of $k$
such that it has the desired {\it initial} spectrum,
apply the projection operator $P_{ij}=\delta_{ij}-\hat k_i\hat k_j$
(to ensure a divergence free magnetic field),
\begin{equation}
B_i({\kk})=B_\star\left[P_{ij}(\kk)-\ii\sigmaM\epsilon_{ijl}
\hat{k}_l\right] g_j({\kk})\, S(k),
\label{Bikk}
\end{equation}
where $g_j(\kk)$ is the Fourier transform of a $\delta$-correlated vector field in three dimensions with Gaussian fluctuations, i.e.,
$g_i({\xx})g_j({\xx'})=\delta_{ij}\,\delta^3(\xx-\xx')$,
$k_0$ is the initial wavenumber of the energy-carrying eddies
and $S(k)$ determines the spectral shape with
\begin{equation}
S(k)={k_0^{-3/2} (k/k_0)^{\alpha/2-1} \,\exp[-{\mathcal G}\,(k^2/k_0^2-1)]
\over[1+(k/k_0)^{2(\alpha+5/3)}]^{1/4}},
\label{Sfunction}
\end{equation}
where ${\mathcal G}=0$ in most cases, and ${\mathcal G}=1$ in some special cases
where the initial power is more strongly concentrated around $k=k_0$.
This results in a random magnetic field with the desired magnetic energy
and helicity spectra and obeys
\begin{equation}
{k\HM(k,t_\star)\over2\EM(k,t_\star)}= {2\sigmaM\over1+\sigmaM^2}\equiv\epsM.
\end{equation}
A similar scheme allows us to generate the velocity field,
\begin{equation}
u_i({\kk})=u_\star\left[P_{ij}(\kk)-\ii\sigmaK\epsilon_{ijl} \hat{k}_l\right] g_j({\kk})\, S(k).
\end{equation}
These initial condition are readily implemented as part of the {\sc Pencil Code}.

We now consider possible initial conditions in a cosmological scenario, where we
have in mind magnetic fields generated at the electroweak phase transition.
In the standard model, the electroweak phase transition is of second order and
CP violation is very weak. However, we also know that the standard model is
incomplete, most convincingly because of the observed non-vanishing neutrino
masses. In addition, the standard model does not contain a candidate for
cosmological dark matter. Neither does it successfully explain the observed
baryon asymmetry of the universe. Hence it is almost certain that there is
fundamental physics beyond the standard model.

The exact nature of what
lies beyond the standard model is unclear. Yet we expect beyond-standard-model
(BSM) physics to explain neutrino masses and contain a suitable dark matter candidate
and also have a successful baryogenesis mechanism. The requirement of
baryogenesis points to some general features essential to BSM as first outlined
by Sakharov~\cite{Sakharov:1967dj}: the model should have strong departures from thermal
equilibrium and should contain significant violations of charge conjugation (C) symmetry,
CP conjugation symmetry, and baryon number.

In the present context,
it is possible that strong departures from thermal equilibrium might occur during
strong first order phase transitions, in which case the cosmological medium
could become turbulent. Thus we would like to include fluid kinetic energy as
an initial condition. Electroweak symmetry breaking also leads to the
production of magnetic fields~\cite{Vachaspati:1991nm}. In addition, baryon number violating
processes lead to the generation of helical magnetic fields~\cite{Vachaspati:2001nb,Copi:2008he}.
If there is significant violation of C and CP, helicity might be large. One may
also expect C and CP violation to leak into the kinetic motion, in which case
the initial conditions would have non-vanishing kinetic helicity.

To keep the discussion as general as possible we consider three different cases
for the initial conditions: (i) magnetically dominant turbulence, (ii) kinetically dominant
turbulence, and (iii) equipartition between magnetic and kinetic energy densities.
In every case, there are several parameters that we have to choose
that quantify the magnetic and kinetic, energy and helicity spectra,
such as $B_\star$,\footnote{Equivalently we can use $b_\star$, see
Eq.~(\ref{bstar}).
Note $b_\star =1$ corresponds the case with the maximal value of the
magnetic field strength allowed by BBN} $u_\star$, $\sigmaK$, and
$\sigmaM$ that we defined in Sec.~\ref{subsec:initialconditions}.
In addition, it is assumed
that the phase transition leads to a peak in the spectra at some fraction, $\gamma_\star$, of the Hubble scale. For example, $\gamma_\star$ will depend on the bubble size at percolation
in the case of a first order phase transition.
The resulting magnetic field values are given
for several choices of
the parameters.

An important control parameter is the initial ratio of the normalized
rms magnetic field (or Alfv\'en velocity) and rms velocity defined as
\begin{equation}
Q_\star=B_\star/(\rho_\star^{1/2} u_\star).
\end{equation}
In this work, we consider the values 10, 1, and 0.1, corresponding to
magnetically dominant, equipartition, and magnetically subdominant cases.
We also consider the time-dependent quantity $Q(t)=\vA/\urms$, and list,
in particular, the value at the last time, $Q_{\rm e}=Q(t_{\rm e})$.
Furthermore, we quote the Reynolds number, $\Rey=\urms\xiM/\nu$,
at $t=t_{\rm e}$.

\paragraph{Magnetically dominant turbulence.}
\label{sec:init-mag}
For the magnetically dominant turbulence we assume that the velocity
field is small initially.
The magnetic field spectrum must satisfy the causality requirements,
i.e., the magnetic field two point correlation function $\langle B_i({\bf
x}) B_j({\bf x}+{\bf r}) \rangle \equiv {\mathcal B}_{ij}(r) \rightarrow
0$ for $r \geq \xiM$ where $\xiM$ is the magnetic field correlation length
scale with the maximal value defined through the comoving Hubble horizon
radius, and we have used the isotropy condition, ${\mathcal B}_{ij}({\bf
r}) = {\mathcal B}_{ij}(|{\bf r}|)$.
The causality condition requires that
$\EM(k,t_\star) \propto k^\alpha$ for $k \rightarrow 0$ together with
the requirement that $F_{ij}({\bf k})$ is analytical for a solenoidal
magnetic field (divergence-less condition $\nab\cdot\BB=0$) this
leads to the Batchelor spectrum $\alpha=4$ \cite{Durrer:2003ja}.

The initial peak position of the magnetic field spectrum is determined
by the phase transition bubble size (i.e., the $\gamma_\star$-parameter).
The ratio between the magnetic and kinetic energy at the initial moment is a large number
$\EEM(t_\star)/\EEK(t_\star)\gg1$, and at any given wavenumber $k$ the
magnetic field spectral energy density is dominant, $\EM(k,t_\star)\gg\EK(k,t_\star)$.
This class of initial conditions is realized in most baryogenesis
mechanisms during cosmological phase transitions.
It can be also applied when the magnetic field was generated at earlier
epochs and undergoes coupling with primordial plasma within the Hubble
horizon.

\paragraph{Kinetically dominant turbulence.}
\label{sec:init-hyd}
In the case of kinetically dominant turbulence, the initial Alfv\'en velocity is
negligibly small compared to the turbulence turnover velocity,
i.e.\ the magnetic field energy density is negligibly small compared to
the kinetic energy density, $\EEM(t_\star) \ll \EEK(t_\star)$, and at a
given wavenumber $k$ the magnetic spectral energy density is subdominant,
$\EM(k,t_\star) \ll \EK(k,t_\star)$.
This class of initial conditions can be realized for the strong
phase transitions when the turbulence turnover velocity
$u_T(t_\star) \simeq 0.3$, that is a consequence of high enough values for $\tilde\alpha$ and $\tilde\kappa$ parameters, and it agrees with the
BBN bound on the relativistic energy density,
see Sec.~\ref{sec:magnetogenesis}.
The initial velocity spectrum can be approximated by the white
noise spectrum $\EK(k,t_\star) \propto k^2$ (that ensures the
causality requirement) or by the Batchelor spectrum $\EK(k,t_\star)
\propto k^4$ (that ensures the causality and divergenceless
requirements).
Interestingly in the latter case the initially solenoidal velocity field acquires
a longitudinal structure through the interaction with the magnetic field,
see below.
In addition we study the evolution of a magnetic field that initially has a white
noise spectrum.

\paragraph{The Case of Equipartition.}
\label{sec:init-equi}
The case of equipartition between magnetic and kinetic spectral energy
densities $\EEM(t_\star) \simeq \EEK(t_\star)$
is hard to be realized in the early universe and requires very specific
physical conditions during phase transitions.
We study this case for completeness.

\subsection{Simulation parameters and analysis tools}
\label{simparams}

We compute magnetic and kinetic energy spectra, $\EM(k,t)$ and $\EK(k,t)$,
respectively, and evaluate corresponding magnetic and kinetic correlation
lengths using \Eq{correlation-length}.
We define a time-dependent Reynolds number, $\Rey=\urms\xiM/\nu$, and
quote approximate values characteristic of the late time evolution.

\begin{figure*}[t!]\begin{center}
\includegraphics[width=\textwidth]{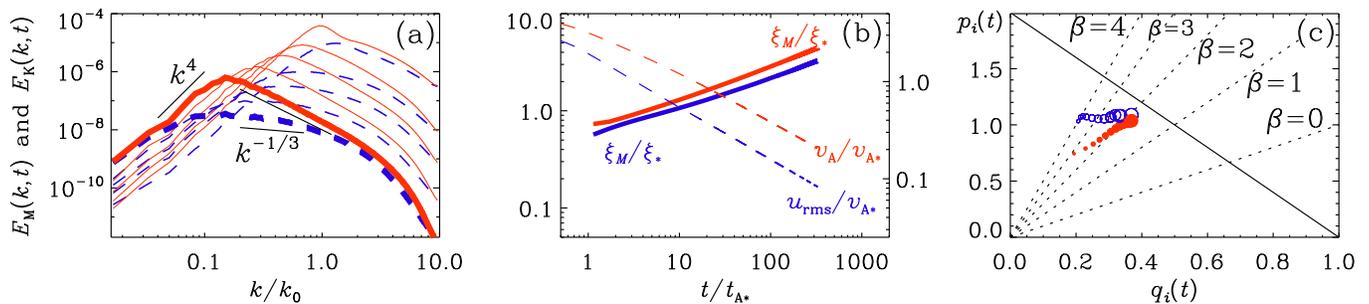}
\end{center}\caption[]{
Run~A with $Q_\star=10$, $\Rey=130$, Batchelor spectrum $\alpha=4$,
so $\nu_\star=10^{-5}\cs/k_1$ and $r=-0.43$ are used,
and no helicity is applied, i.e., $\sigmaM=\sigmaK=0$.
(a) $\EM$ (red, solid) and $\EK$ (blue, dashed)
at times $t/\tauAz=4$, 30, 120, 500, and 2000.
The last time is indicated by thick lines.
(b) $\xiM/\xiz$ (red, thick) and $\xiK/\xiz$ (blue, thick) with scale
on the left, together with $\vA/\vAz$ (red, dashed) and $\urms/\vAz$
(blue, dashed) with scale on the right.
(c) $pq$ diagram showing the evolution of $\PPM$ (red, filled symbols)
and $\PPK$ (blue, open symbols).
The symbol size increases with time.
The equilibrium line $p=2(1-q)$ is shown as solid,
while the $\beta=\const$ lines are dotted.
}\label{pkt1152_KH1152tnuk4b_sig0_M10}\end{figure*}

As demonstrated earlier \citep{Brandenburg:2016odr}, $E_i(k,t)$
with $i={\rm M}$ and ${\rm K}$ can be collapsed onto a function
$\phi_i(\kappa)$ of a single argument $\kappa=k\xi_i(t)$ via
\begin{equation}
E_i(k,t)=\xi_i^{-\beta_i}\phi_i(k\xi_i),
\label{collapse2}
\end{equation}
where $\beta_i$ quantifies the decay of the spectral energy around the
wave number $k=\xi_i^{-1}$, which itself decreases approximately
like a power law with $\xi_i(t)\propto t^{q_i}$, where $q_i$ is
a scaling exponent.
Since ${\cal E}_i(t)=\int E_i(k,t)\,dk$, it also decays like a
power law with ${\cal E}_i(t)\propto t^{-p_i}$, where
\begin{equation}
p_i=(\beta_i+1)\,q_i.
\label{pbetaq}
\end{equation}
The values of $\beta_i$ are believed to depend on the physics
that governs a particular case \cite{Brandenburg:2016odr}.

It is convenient to define and plot instantaneous scaling exponents as
$p_i(t)=d\ln{\cal E}_i/d t$ versus $q_i(t)=d\ln\xi_i/d t$ for
$i={\rm M}$ and ${\rm K}$ and discuss the evolution of the point
\begin{equation}
\PP_i=(p_i,q_i)
\end{equation}
in the $pq$ diagram. Solutions that obey invariance under rescaling \cite{Ole97,Brandenburg:2016odr},
\begin{equation}
k\to k'\ell\quad\mbox{and}\quad
t\to t'\ell^{1/q_i},
\end{equation}
all lie on the line $p_i=2(1-q_i)$ in this diagram.
The functions $\phi_i(\kappa)$ are universal
functions for given $\beta_i$ and thus $q_i$.
If that is the case, then $q_i=2/(\beta_i+3)$.

We are particularly interested in the possibility of an inverse cascade,
which means that the magnetic energy increases at small wavenumbers,
even though the total energy decreases.
This implies that
\EQ
s_i\equiv\partial\ln E_i(k,t)/\partial\ln t >0
\quad\mbox{for $k\ll\xi_i(t)^{-1}$}.
\EN
At small $\kappa=k\xi_i(t)$,
we have $\phi_i(\kappa)\propto\kappa^\alpha$,
and therefore
\EQ
s_i=(\alpha_i-\beta_i)\,q_i,
\label{sRelation}
\EN
which implies that large initial slopes (e.g., $\alpha=4$) and small
values of $\beta$, e.g., when the decay is governed by the conservation
of magnetic helicity ($\beta=0$) or the mean squared vector potential
($\beta=1$) will lead to an inverse cascade, but not when $\beta\ge2$)
\cite{Brandenburg:2016odr}.

\begin{figure*}[t!]\begin{center}
\includegraphics[width=\textwidth]{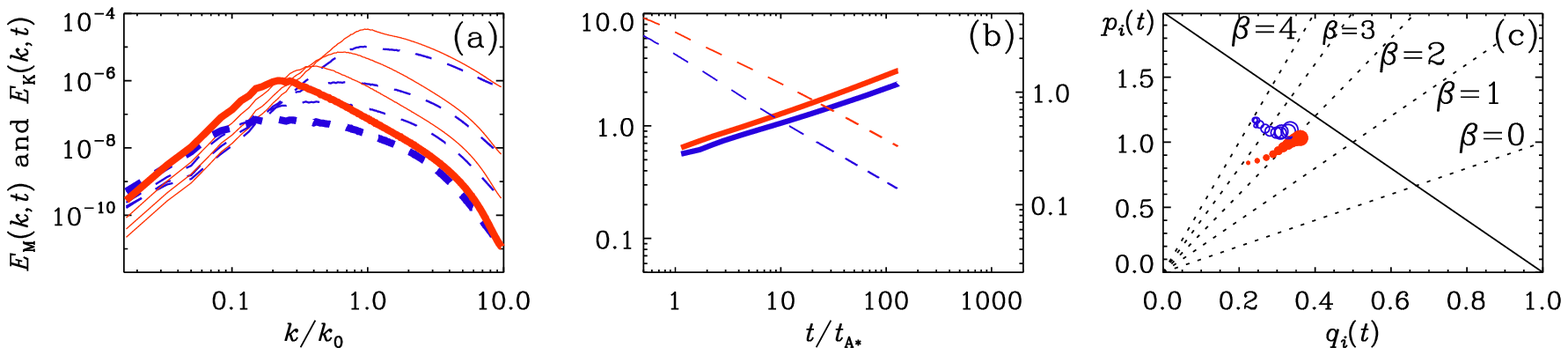}
\end{center}\caption[]{
Same as \Fig{pkt1152_KH1152tnuk4b_sig0_M10},
but for Run~B with $Q_\star=1$ and $\Rey=100$.
The times in (a) are $t/\tauAz=4$, 40, 180, and 800.
}\label{pkt1152_KH1152tnuk4b_sig0}\end{figure*}

\begin{figure*}[t!]\begin{center}
\includegraphics[width=\textwidth]{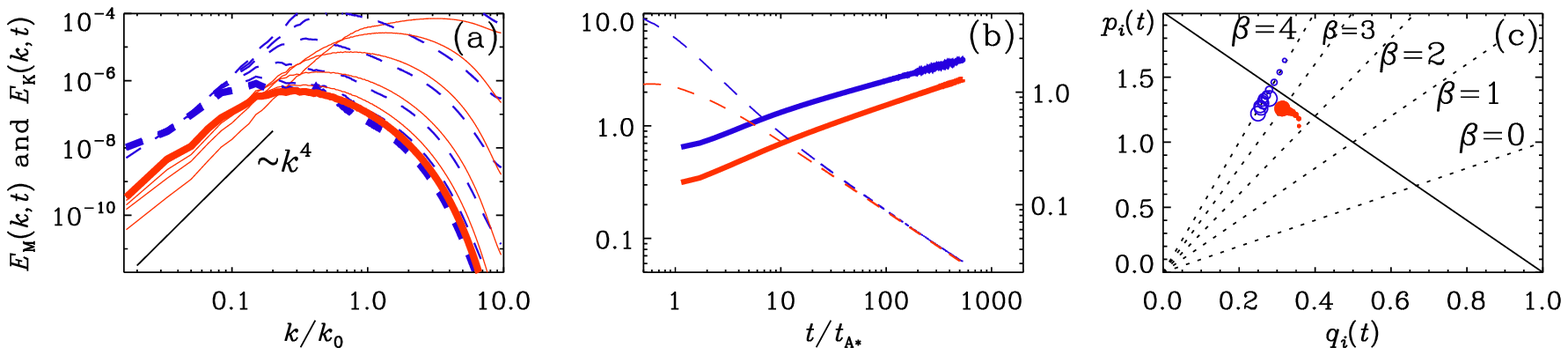}
\end{center}\caption[]{
Same as \Fig{pkt1152_KH1152tnuk4b_sig0_M10},
but for Run~C with $Q_\star=0.1$ and $\Rey=35$.
The times in (a) are $t/\tauAz=0.4$, 4, 18, 80, and 200.
}\label{pkt1152_KH1152tnuk4b_sig0_M01}\end{figure*}

\section{Results}
\label{sec:results}

\subsection{Batchelor spectrum and no helicity}

\begin{table}[b!]\caption{Summary of our runs
}\vspace{12pt}\centerline{\begin{tabular}{cccccccccccc}
Run & $\sigmaK$ & $\sigmaM$ & $\alpha$ & ${\cal G}$ &
$Q_\star$ & $Q_{\rm e}$ & $t_{\rm e}/\tauAz$ & $\betaM$ &
$q_{\rm M}$ & $p_{\rm M}$ & Figure\tablenote{
BKT refers to the nonhelical run of Ref.~\cite{BKT15}
and BK to a fully helical of Ref.~\cite{Brandenburg:2016odr}.
$Q_\star$ and $Q_{\rm e}$ refer to the values of $Q$ at the
beginning and end of the run, respectively.
The instantaneous scaling exponents $\betaM$,
$q_{\rm M}$, and $p_{\rm M}$ are given at the end of the run,
whose normalized time $t_{\rm e}/\tauAz$ is given.
}
\\
\hline
BKT&0 & 0   & 4 &0&$\infty$& 2.8 & 1335 & 1.2 & 0.47 & 1.02 & Ref.~\cite{BKT15}\\ %
A & 0 & 0    & 4 & 0 & 10  & 2.5 &  206 & 1.8 & 0.37 & 1.04 & \Fig{pkt1152_KH1152tnuk4b_sig0_M10} \\
B & 0 & 0    & 4 & 0 &  1  & 2.4 &  114 & 1.9 & 0.36 & 1.03 & \Fig{pkt1152_KH1152tnuk4b_sig0} \\
C & 0 & 0    & 4 & 0 & 0.1 & 1.0 &  460 & 3.0 & 0.31 & 1.26 & \Fig{pkt1152_KH1152tnuk4b_sig0_M01} \\
D & 0 & 0    & 2 & 0 &  1  & 3.2 &  208 & 1.7 & 0.38 & 1.03 & \Fig{pkt1152_KH1152tnuk2b_sig0} \\
E & 0 & 0    & 2 & 1 &  1  & 2.6 &  170 & 1.7 & 0.36 & 0.95 & \Fig{pkt1152_KH1152tnuk2cG60_sig0} \\
F & 0 & 0    & 2&1&$\infty$& 2.6 &  170 & 1.7 & 0.35 & 0.94 & \Fig{pkt1152_KH1152tnuk2dG60_sig0} \\
G & 0 & 0.03 & 2 & 1 &  1  & 3.2 & 1024 & 0.3 & 0.55 & 0.73 & \Fig{pkt1152_KH1152tnuk2cG60_sig003} \\
H & 1 & 1    & 4 & 0 &  1  & 1.3 &  562 & 0.6 & 0.46 & 0.76 & \Fig{pkt1152_KH1152tnuk4b_sig1_M01} \\
I & 1 & 0    & 4 & 0 &  1  & 2.3 & 1250 & 0.2 & 0.49 & 0.58 & \Fig{pkt1152_KH1152tnuk4b_sig1b_M01} \\
J & 1 &$-1$  & 4 & 0 &  1  & 2.9 &  460 & 0.1 & 0.48 & 0.57 & \Fig{pkt1152_KH1152tnuk4b_sig1c_M01} \\
BK & 0 & 1   & 4 &0&$\infty$&4.2 & 1025 & 0.0 & 0.59 & 0.62 & Ref.~\cite{Brandenburg:2016odr}\\
\end{tabular}}
\label{TSummary}
\end{table}

We begin by comparing the evolution of initially nonhelical velocity and
magnetic fields for $Q_\star=10$, 1, and 0.1, responding to Runs~A--C;
see \Figss{pkt1152_KH1152tnuk4b_sig0_M10}{pkt1152_KH1152tnuk4b_sig0_M01}
and \Tab{TSummary}.
In all three cases, we plot $E_i(k,t)$ at selected times, normalized
by the initial Alfv\'en time $\tauAz$.
We also
show the evolution of ${\cal E}_i(t)$ and $\xi_i$, as well as a parametric
representation of the instantaneous scaling exponents $p_i(t)$ versus
$q_i(t)$ for $i={\rm M}$ and ${\rm K}$ ($pq$ diagram).
In all three cases (Runs~A, B, and C), there is inverse energy transfer
at small $k$, which is in agreement with \Eq{sRelation}.

Remarkably, Runs~A and B are rather similar at later times, i.e., for $t/\tauAz\ga10$,
where $Q(t)\approx10$, which agrees with the initial value $Q_\star=10$ for Run~A, but not with that Run~B,
where the initial ratio was unity. The resulting values of $\pM\approx1$ are similar to those obtained
earlier from an initial condition obtained from a run that was driven
for a short time with a magnetic forcing function \cite{BKT15},
which is marked in \Tab{TSummary} by BKT, where $\beta$ turned out
to be close to 1 instead of the present value of 2.
For Run~C, on the other hand, even though $Q$ was initially 0.1, it
reaches unity at later times; see \Fig{pkt1152_KH1152tnuk4b_sig0_M01}.

\begin{figure*}[t!]\begin{center}
\includegraphics[width=\textwidth]{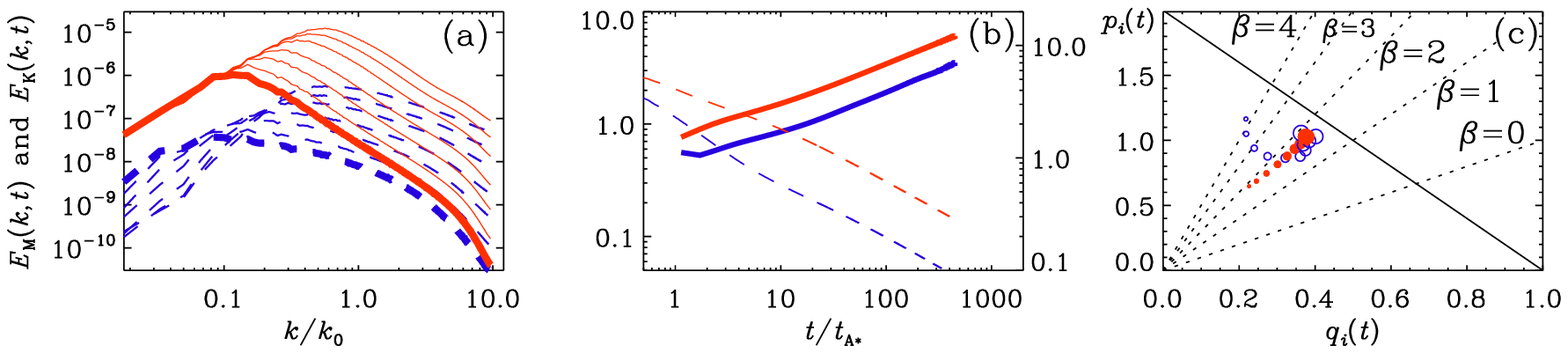}
\end{center}\caption[]{
Similar to \Fig{pkt1152_KH1152tnuk4b_sig0_M10},
but for Run~D with $Q_\star=1$, $\Rey=600$, and
$\alpha=2$, so $\nu_\star=10^{-6}\cs/k_1$ and $r=-0.20$ are used.
The times in (a) are $t/\tauAz=40$, 80, 150, 400, 800, 1600, and 3000.
}\label{pkt1152_KH1152tnuk2b_sig0}\end{figure*}

Indeed, comparing the $pq$ diagrams for all three cases, we see again that
for Runs~A and B, both $\PPM$ and $\PPK$ evolve along the $\beta=2$ line
toward the equilibrium line where $p=2(1-q)$ and thus $p=6/5$ and $q=2/5$.
By contrast, for Run~C, the $\PP_i$ (with $i={\rm M}$ and ${\rm K}$) evolve
towards the $\beta=4$ line, although during the time interval of the
run shown here.
Furthermore, the $\PP_i$ seem to move away from the equilibrium line. At present, we do not know whether this could be
an artifact of limited scale separation
($k/k_1$ is too small) toward small values of $k$
and also of the limited inertial range between
$k_0$ and $k_{\rm D}$ above which the spectra stop being power laws.

\subsection{White-noise spectrum and no helicity}

Let us now turn to simulations with $\alpha=2$, which was recently studied
in Ref.~\cite{Reppin:2017uud}, where it was found that no inverse transfer
occurs in that case.
Here we also compare with simulations where an additional Gaussian
profile is included in the
initial spectrum (${\cal G}=1$); see \Eq{Sfunction}.
We only consider cases where $Q_\star=1$ or $\to\infty$.

\begin{figure*}[t!]\begin{center}
\includegraphics[width=\textwidth]{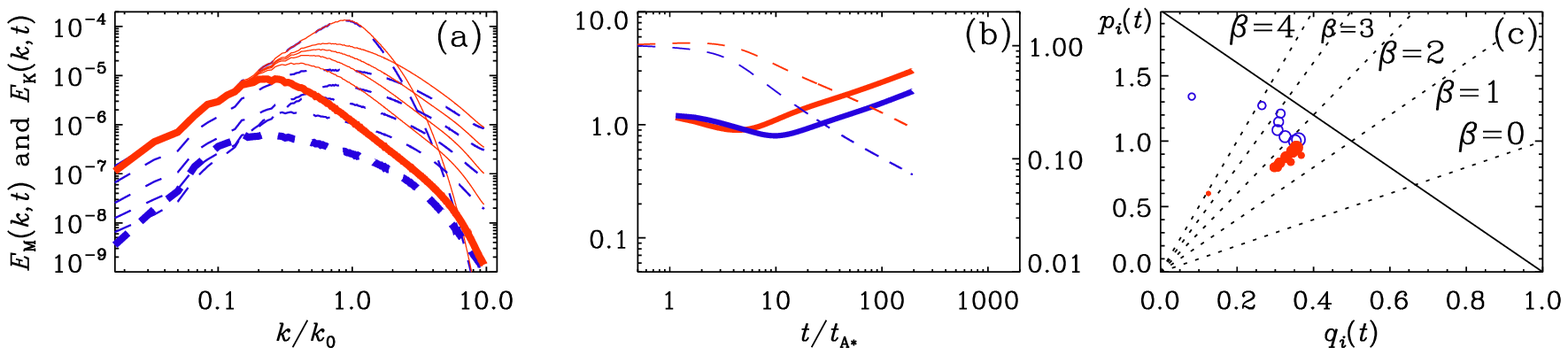}
\end{center}\caption[]{
Same as \Fig{pkt1152_KH1152tnuk2b_sig0},
but for Run~E with ${\cal G}=1$ and $\Rey=200$.
The times in (a) are $t/\tauAz=0.6$, 6, 12, 20, 50, and 200.
}\label{pkt1152_KH1152tnuk2cG60_sig0}\end{figure*}

\begin{figure*}[t!]\begin{center}
\includegraphics[width=\textwidth]{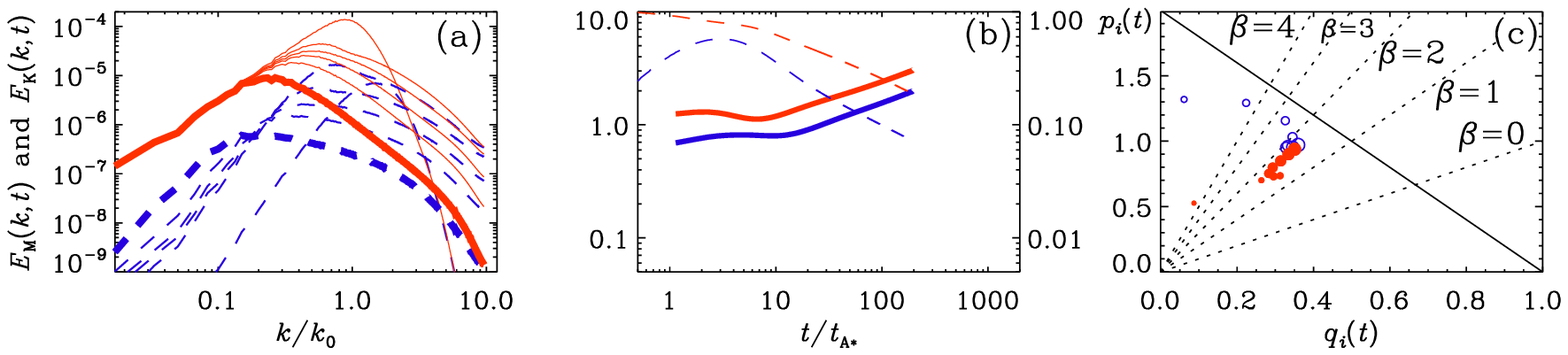}
\end{center}\caption[]{
Same as \Fig{pkt1152_KH1152tnuk2cG60_sig0}, but for Run~F with $Q_\star\to\infty$,
i.e., $u_\star=0$, and $\Rey=200$.
The times in (a) are $t/\tauAz=0.6$, 6, 12, 20, 50, and 200.
}\label{pkt1152_KH1152tnuk2dG60_sig0}\end{figure*}

Not surprisingly, the cases with ${\cal G}=0$ (Run~D; see
\Fig{pkt1152_KH1152tnuk2b_sig0}) and ${\cal G}=1$ (Run~D; see
\Fig{pkt1152_KH1152tnuk2cG60_sig0}) are rather similar, except that
the early time evolution is closer to equipartition
We also compare with the case $Q_\star\to\infty$ (Run~F).
Again, it has the same late-time evolution as Runs~D and E,
but the early time evolution is now close to that of Run~D; see
\Fig{pkt1152_KH1152tnuk2dG60_sig0}.

In all these cases, $\PP$ evolves along the $\beta=2$ line towards
the equilibrium line.
This implies that in these cases there is no inverse transfer,
see \Eq{sRelation}, which is consistent with Ref.~\cite{Reppin:2017uud}.
As we already noted, the white noise spectrum for the initial magnetic
field has an academic interest only because we expect causality to limit
the power on large length scales to sub-white noise levels; see footnote~2.

\subsection{White-noise spectrum with magnetic helicity}

The case of fractional helicity has been studied previously
\cite{Tevzadze:2012kk} in connection with QCD phase transition-created
initial magnetic fields.
In their studies, $\alpha=4$ was used, but the resolution was only $512^3$.

\begin{figure*}[t!]\begin{center}
\includegraphics[width=\textwidth]{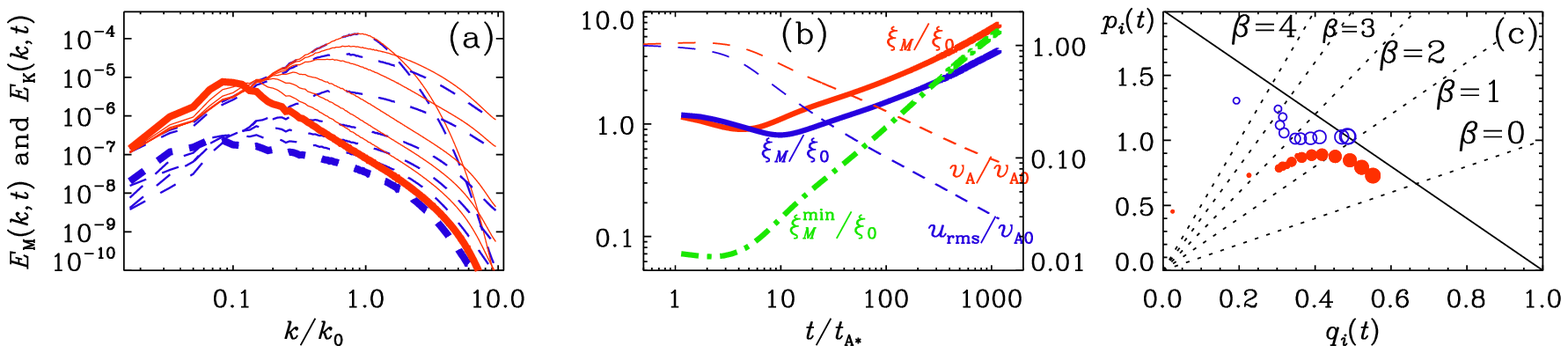}
\end{center}\caption[]{
Similar to \Fig{pkt1152_KH1152tnuk2cG60_sig0}, but for Run~G with $\sigmaM=0.03$
and $\Rey=300$.
The times in (a) are $t/\tauAz=0.6$, 4, 18, 120, 300, 600, and 1200.
In (b), the evolution of $\xiM^{\min}$ is shown as a green dashed-dotted line.
}\label{pkt1152_KH1152tnuk2cG60_sig003}
\end{figure*}

We now discuss the case with $\alpha=2$ (Run G).
In contrast to the earlier case with $\alpha=4$ \cite{Tevzadze:2012kk},
there is now no inverse transfer at early times when the
magnetic energy is still strong.
As in earlier work, we plot the evolution of $\xiM$,
as defined in Eq.~(\ref{correlation-length}),
which increases like $t^{1/2}$.
We compare this with $\xi_M^{\min}$, defined in Eq.~(\ref{xiMdefn}), which
increases with time since $\HHM(t)=\const$ and $\EEM\propto t^{-1}$.
The result is shown in \Fig{pkt1152_KH1152tnuk2cG60_sig003}.
Evidently, $\xiM^{\min}(t)\propto t$, and so $\xiM^{\min}(t)$ will be
equal to $\xiM$ after some time.
The initial value of $\xiM^{\min}(t)$ depends on the fractional helicity
and is given by $\epsM k_0^{-1}$.
It turns out that the late-time sub-inertial spectrum for the magnetic
field changes from a $k^2$ (white noise spectrum) to a $k^4$
(Batchelor spectrum) at the time when the magnetic field begins
to be fully helical.

\begin{figure*}[t!]\begin{center}
\includegraphics[width=\textwidth]{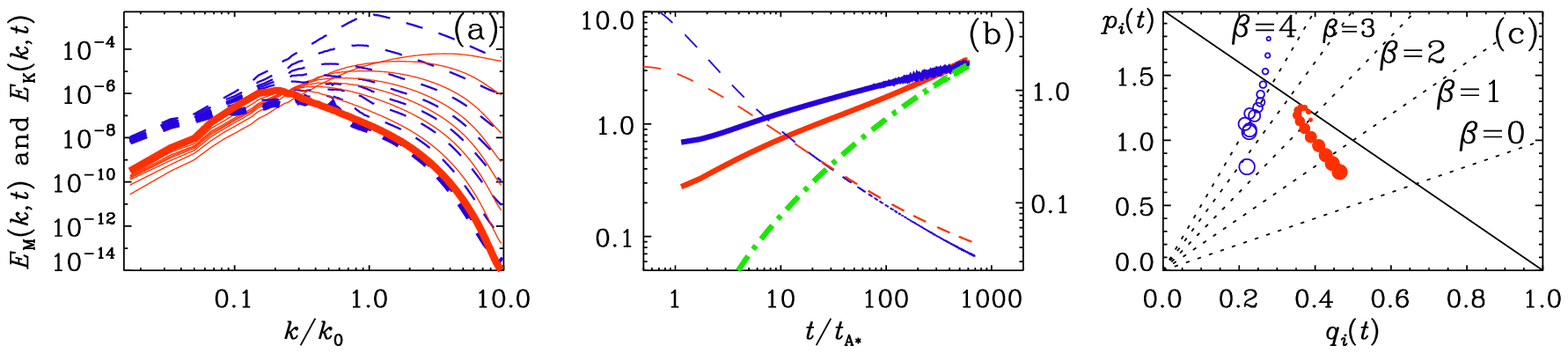}
\end{center}\caption[]{
Similar to \Fig{pkt1152_KH1152tnuk4b_sig0}, but for Run~H with $\sigmaK=\sigmaM=1$
and $\Rey=65$.
The times in (a) are $t/\tauA=0.5$, 3, 10, 25, 50, 100, 250, and 500.
}\label{pkt1152_KH1152tnuk4b_sig1_M01}\end{figure*}

\begin{figure*}[t!]\begin{center}
\includegraphics[width=\textwidth]{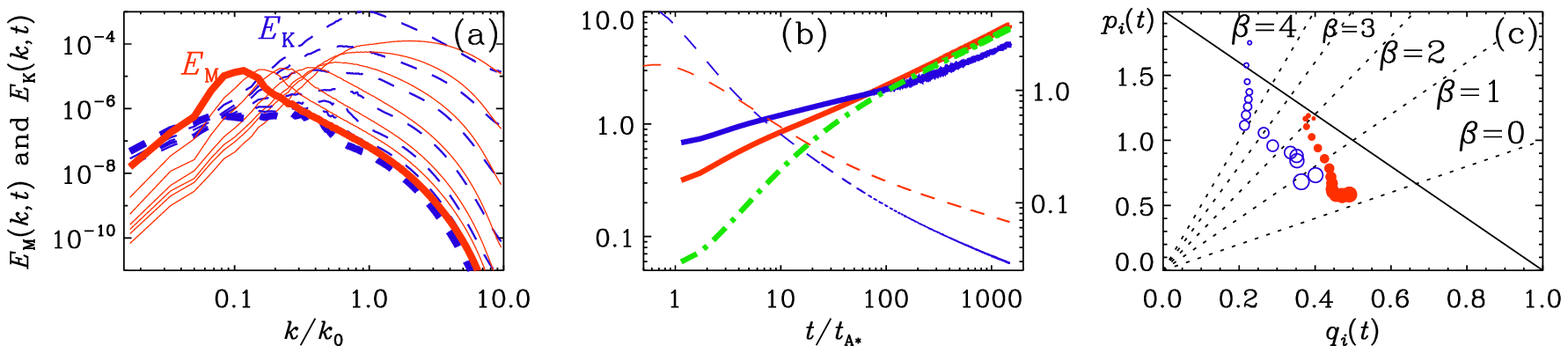}
\end{center}\caption[]{
Similar to \Fig{pkt1152_KH1152tnuk4b_sig1_M01},
but for Run~I with $\sigmaK=1$, $\sigmaM=0$ and $\Rey=160$.
The times in (a) are $t/\tauA=1$, 4, 14, 60, 180, and 600.
}\label{pkt1152_KH1152tnuk4b_sig1b_M01}\end{figure*}

\begin{figure*}[t!]\begin{center}
\includegraphics[width=\textwidth]{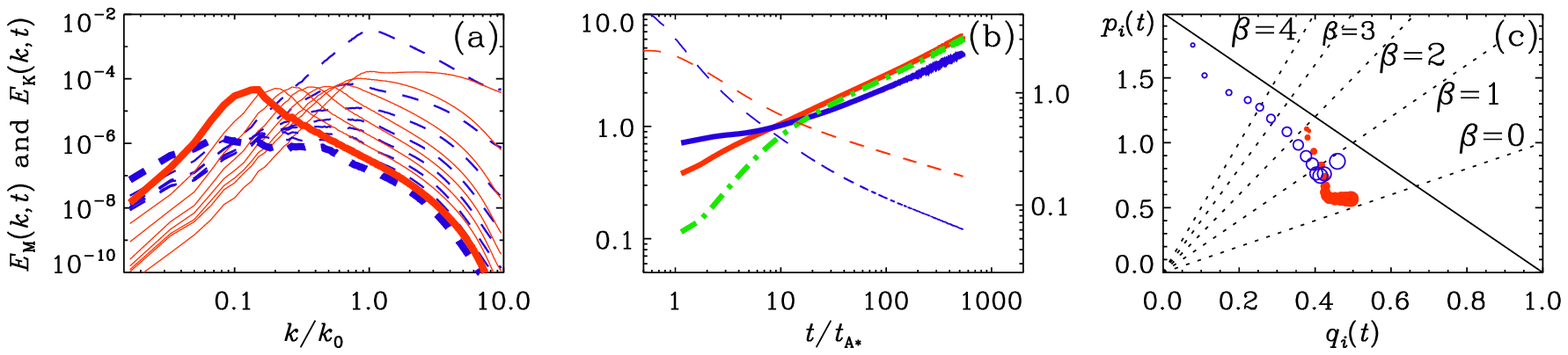}
\end{center}\caption[]{
Similar to \Fig{pkt1152_KH1152tnuk4b_sig1_M01},
but for Run~J with $\sigmaK=1$, $\sigmaM=-1$ and $\Rey=160$.
The times in (a) are $t/\tauA=0.5$, 3, 10, 25, 50, 100, 250, and 500.
}\label{pkt1152_KH1152tnuk4b_sig1c_M01}\end{figure*}

\subsection{Batchelor spectrum with initial kinetic helicity}

The initial presence of kinetic helicity has profound effects on the
evolution of the magnetic field.
Kinetic helicity leads to an $\alpha$ effect, i.e., the
destabilization of a large-scale magnetic field.
The details of this process in decaying turbulence were studied
in Ref.~\cite{Brandenburg:2017rnt}, where it was found that the initial
kinetic helicity gets transformed efficiently into magnetic helicity such
that the residual helicity, $\bra{\oo\cdot\uu}-\bra{\JJ\cdot\BB}/\rho_\star$
is approximately constant.
During the time of their runs, the magnetic helicity $\bra{\AAA\cdot\BB}$
was still increasing, so one expects to reach the familiar behavior with
$p_i=q_i=2/3$ at much later times.

In \Tab{TSummary}, the corresponding results for $p$ and $q$ from
Ref.~\cite{Brandenburg:2016odr} are marked with BK.
$\PP_i$ evolves towards the $\beta=0$ line, but it is
still far away from the ultimate equilibrium line $p=2(1-q)$.
Instead, we see that in \Figss{pkt1152_KH1152tnuk4b_sig1_M01}
{pkt1152_KH1152tnuk4b_sig1c_M01}, $q_{\rm M}=0.4$--$0.5$ during an
extended time interval, and that $p_{\rm M}=0.5$--$0.6$, while in
the equilibrium state we would expect $p_{\rm M}=1.2$--$1.0$.

\subsection{Comparison with the equilibrium line}

\begin{table}[b!]\caption{
Scaling exponents and relation to physical invariants and their dimensions.
}\vspace{12pt}\centerline{\begin{tabular}{rrrll}
$\beta$ & $q\quad\quad\;$ & $p\quad\quad$ & $\quad$inv. & dim. \\
\hline
4 & $\quad2/7\approx0.286$ & $\quad10/7\approx1.43$ & $\quad{\cal L}$ & $[x]^7[t]^{-2}$ \\
2 & $2/5=0.400$            &   $6/5=1.20$           & $\quad{\cal S}$ & $[x]^5[t]^{-2}$ \\
1 & $2/4=0.500$            &   $4/4=1.00$           & $\quad\bra{\AAA_{\rm 2D}^2}$ & $[x]^4[t]^{-2}$ \\
0 & $2/3\approx0.667$      &   $2/3\approx0.67$     & $\quad\bra{\AAA\cdot\BB}$  & $[x]^3[t]^{-2}$   \\
\label{TSum}\end{tabular}}
\end{table}

In \Tab{TSum}, we summarize the anticipated values of $q$ and $p$ that
would be expected for given values of $q$ or $\beta$ if the solution
were to lie on the equilibrium line in the $pq$ diagram.
These different cases are based on the dimensions of potentially
conserved quantities such as the Loitsiansky and Saffman integrals,
\be
{\cal L}=\int\rr^2\bra{\uu(\xx)\cdot\uu(\xx+\rr)}\,d\rr\propto\ell^5 u_\ell^2
\label{loitsiansky}
\ee
and
\be
{\cal S}=\int\bra{\uu(\xx)\cdot\uu(\xx+\rr)}\,d\rr\propto\ell^3 u_\ell^2,
\label{saffman}
\ee
respectively \cite{Dav10}, with typical velocity $u_\ell$ on scale
$\ell$, the conservation of magnetic helicity, $\bra{\AAA\cdot\BB}$,
and the possible conservation of the mean squared vector potential,
$\bra{\AAA^2}$, which is known to be conserved in two-dimensions (2D).

Comparing with the numerical results given in \Tab{TSummary}, we see
that for the runs with fractional magnetic helicity or with initial
kinetic helicity, there is a tendency to develop maximal magnetic
helicity at later times.
As a consequence, all those runs are seen to develop toward the $\beta=0$ line.
However, in none of those runs there is a perfect convergence toward
the equilibrium point with $p=q=2/3$, as would be expected in the fully
helical case.
Instead, we find that $q\approx0.5$ and $p\approx0.6$, so the decay is
even slower than with maximum helicity.

The departure from the expected equilibrium position may well be a finite
size effect of the computational domain.
Ideally, one would like to have a much larger numerical resolution,
so as to be able to follow an unimpeded development of the inverse
cascade for both $\EM$ and $\EK$ toward smaller wave numbers.
At the same time, of course, it is important to include large enough
wave numbers to resolve the turbulent inertial and dissipative subranges.

In most of the runs without kinetic or magnetic helicity, the
final values of $q$ are in the range $0.3$--$0.4$, which is again smaller than
what is expected for the equilibrium points $(p,q)=(0.5,1)$, when $\beta=1$
or $(0.4,1.2)$, when $\beta=2$; see \Tab{TSum}.
In those cases, on the other hand, there is a clear trend that $(p,q)$
evolves along the $\beta=2$ line towards the equilibrium point; see
\Figs{pkt1152_KH1152tnuk4b_sig0_M10}{pkt1152_KH1152tnuk4b_sig0}
for $\alpha=4$ and
\Figss{pkt1152_KH1152tnuk2b_sig0}{pkt1152_KH1152tnuk2dG60_sig0} for
$\alpha=2$.

Interestingly, the two groups of runs for $\alpha=4$ and $\alpha=2$
show the same convergence properties along the $\beta=2$ line toward
the equilibrium point $(p,q)=(0.4,1.2)$.
This decay law is suggestive of the case where the Saffman integral
(\ref{saffman}) is conserved.
Thus, we have here is a clear example where the temporal evolutions of
$\EEM$ and $\xiM$ are clearly independent of the initial slope $\alpha$,
where the case with $\alpha=4$ shows inverse cascading while that
with $\alpha=2$ does not, as expected based on \Eq{sRelation}.

The subequipartition case with $Q_\star=0.1$ is different again; see
\Fig{pkt1152_KH1152tnuk4b_sig0_M01}, where we observe a clear development
along the $\beta=4$ line toward the equilibrium point on which the
Loitsiansky integral (\ref{loitsiansky}) is expected to be conserved.

\section{Discussion}

This work has exposed several unknown behaviors of decaying
MHD turbulence.
Firstly, for nonhelical turbulence with an $\alpha=4$ Batchelor spectrum, large initial values of $Q_\star$ (here $Q_\star=1$ and 10) lead to
distinctly different behaviors than small values (here $Q_\star=0.1$). While the former case yields $Q_{\rm e}\equiv Q(t_{\rm e})\approx3$
at the end of our runs (at $t=t_{\rm e}$), the latter case yields
$Q(t_{\rm e})\approx1$; see Run~C in \Tab{TSummary}.
There is at present no indication that all these cases yield ultimately
the same late-time behavior.
However, we cannot exclude the possibility that large and small initial
$Q_\star$ values yield ultimately the same final $Q_{\rm e}$ value.

Second, in the case with $\alpha=2$, no inverse transfer was found
to be possible.
This is because that case also yields $\beta=2$, and so $\alpha=\beta$,
which implies that no inverse transfer is possible; see \Eq{sRelation}.
This is compatible with recent work by Reppin \& Banerjee
\cite{Reppin:2017uud}.

Third, in the case with initial kinetic helicity, a non scale-invariant
behavior is found during an extended period of time where the points
$\PPK$ and $\PPM$ evolve away from the equilibrium line, $p=2(1-q)$.

In view of the early universe, an important lesson is the fact that even
just a small amount of magnetic or kinetic helicity yields the standard
fully helical inverse transfer after a certain time.
The situation is similar in the case where there is only kinetic helicity
initially.
In both cases, $\betaM\approx0$, which implies that $p_{\rm M}=q_{\rm M}$;
see \Eq{pbetaq}.
This also means that $\Brms\propto\xiM^{-1/2}$.
However, unlike the case with initial magnetic helicity where
$p_{\rm M}=q_{\rm M}=2/3$, we find here $p_{\rm M}\approx q_{\rm M}\approx1/2$
during an extended period of time; see
\Figs{pkt1152_KH1152tnuk4b_sig1b_M01}{pkt1152_KH1152tnuk4b_sig1c_M01}.
Ultimately, at very late times, we might still expect
$p_{\rm M}=q_{\rm M}=2/3$, but the time required for this
to happen may be too long.

To put our results into perspective, it is instructive to consider
the evolution of $\Brms$ as a function of $\xiM$, which, in turn, is
a function of time and thus of the scale factor or the inverse
temperature of the universe.
The turbulent evolution of $\Brms$ and $\xiM$ proceeds from the time
of magnetic field generation until recombination.
This implies an increase in the conformal time by twelve orders of
magnitude, and thus eight orders of magnitude in $\xiM\propto t^{2/3}$,
if the initial magnetic field is fully helical.
On the other hand, if there is only initial kinetic helicity, and if
the $\xiM\propto t^{1/2}$ decay law persists for a significant fraction
of time, we might only cover about six orders of magnitude in $\xiM$,
but the field will not decay as much as in the former case.

\begin{figure*}[t!]\begin{center}
\includegraphics[width=\textwidth]{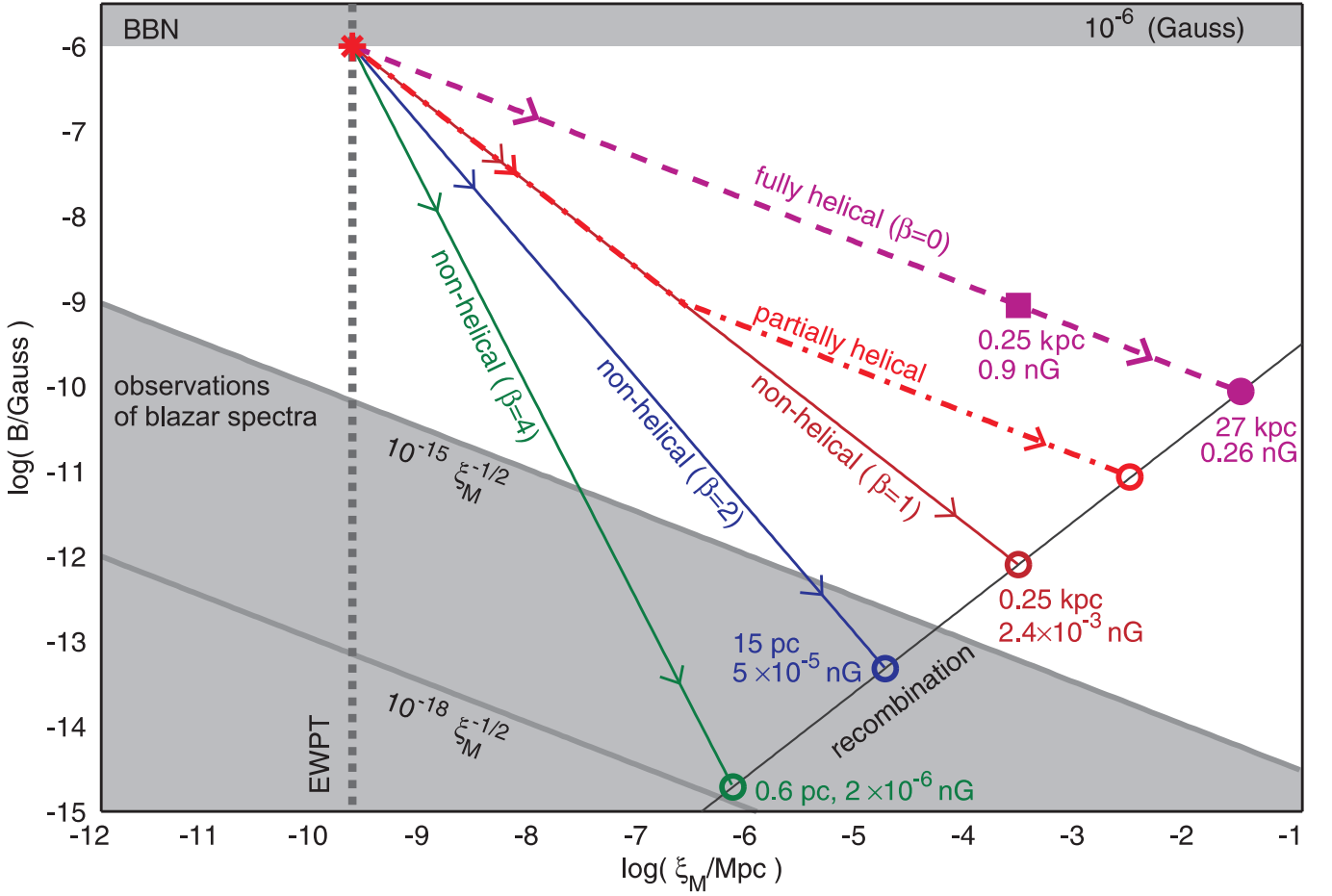}
\end{center}\caption[]{
Turbulent evolution of $\Brms$ and $\xiM$ starting from their
upper limits given by the BBN bound and the horizon scale at the EWPT
for the fully helical case ($\Brms\propto\xiM^{-1/2}$),
the nonhelical case ($\Brms\propto\xiM^{-1}$), and the fractionally
helical case with $\epsMz=10^{-3}$. Circles indicate the final points at recombination for zero or partial
initial magnetic helicity, the filled circle marks the fully helical case,
and the filled square indicates the case with the initial kinetic helicity. The regimes excluded by observations of blazar spectra (upper line:
limits claimed by Neronov and Vovk \cite{Neronov:1900zz}, based on
the consideration of the expected cascade flux in the GeV band produced
by the blazar TeV photons absorbed by the extragalactic background light,
and assuming that the mean blazar TeV flux remains constant; bottom
line corresponds to the limits obtained through
accounting for the fact that the TeV flux activity is limited by the
source observation period (few years) \cite{Taylor:2011bn,Dermer:2010mm}
and BBN limits are marked in gray.
The end of the evolution at recombination is denoted by
the straight line given by the relation in Eq.~(\ref{Brecxirec}), and
the final values of $\Brms$ and $\xiM$ are indicated for
helical and nonhelical scenarios.
}\label{Cons_2}\end{figure*}

Turning now to the cosmological applications of our results,
we are interested in predicting the field characteristics at the
epoch of recombination, $t_{\rm rec}$, for initial conditions specified
at some earlier epoch, $t_\star$.
In Appendix~\ref{characteristics} we show that, if there is sufficient
time for the magnetic field to reach maximal helicity, and if it is not
caused by initial kinetic helicity (which leads to $p\approx q\approx 0.5$
for a long period of time, as in Run~I), then
\begin{equation}
{B_{\rm rec} \over \xi_{\rm rec}} = {B_{\star} \over \xi_{\star}}
\left( {t_{\rm rec} \over t_{\star}} \right)^{-1} ~.
\label{Brecxirec}
\end{equation}
This result is independent of the initial hydromagnetic state and
provides a universal result, applicable to a large number of cases
we have considered.
Note that $t_\star/t_{\rm rec} = T_{\rm rec}/T_\star$.

Let us now discuss the different turbulent decay scenarios for two
cases, the best case scenario where a magnetic field is generated
at the horizon scale with a strength limited by BBN and the second case
where magnetic helicity is generated by the chiral magnetic effect;
see \Fig{Cons_2}.
In the former case, if the initial field is fully helical,
we will reach a magnetic field at a scale of $30\kpc$ with
a strength of $0.3\nG$.
If we only have kinetic helicity initially, and if the
$\xiM\propto t^{1/2}$ behavior persists during the whole time,
we might even get $3\nG$, but only on a scale of $0.3\kpc$.
If the magnetic field stays nonhelical during the entire time,
and if turbulence is magnetically dominated, the field would
again be of a typical scale of about $0.3\kpc$, but now the
field is now significantly weaker -- about $3\times10^{-3}\nG$.
Even magnetic fields amplified by the chiral magnetic effect
cannot have helicity in excess of
$\bra{\BB^2}\,\xiM\approx5\times10^{-38}\G^2\Mpc$ if the chiral
asymmetry is set by the temperature.
This might still be compatible with the most conservative lower limits
on the magnetic field strength derived from blazar spectra,
when accounting that the TeV flux activity is limited by the source
observation period (few years) \cite{Taylor:2011bn,Dermer:2010mm},
but not with stronger fields on large length scales claimed in
Ref. \cite{Neronov:1900zz}, through assumption of the constant mean
blazar TeV flux.

\section{Conclusions}

To understand the evolution of cosmic magnetic fields, we have
considered a broad range of different initial conditions:
magnetically and kinetically dominated cases, with and without
helicity either in the magnetic or the velocity field, as well as
with shallow and steeper initial energy spectra.
Our results are best summarized by presenting them parametrically
in a $\Brms$ versus $\xiM$ diagram.
The resulting trajectories have different slopes,
$-(1+\beta)/2$, and cover different extents in $\Delta\log\xiM
=q\Delta\log t$ in time.
The most shallow slope is $1/2$ in the helical case, where $\beta=0$.
This is independent of whether helicity is initially in the
magnetic field or in the velocity.

Although the two cases are essentially the same as far as the slope
is concerned, there is a difference in terms of the length scales
covered during the evolution.
The largest range of scales is covered when the initial magnetic field
is fully helical and $q=2/3$, while it is only $q=1/2$ when only the
velocity is initially helical.
Consequently, because $p=q$ in the fully helical case, the magnetic
field decays less in the latter case.
However, it is not clear whether there is any physical mechanism
that can create kinetic helicity throughout the entire universe.
Familiar effects in dynamo theory that involve rotation and nonuniformity
always produce positive and negative sign at the same time, so there is
no net effect on larger scales.
For the magnetic field, on the other hand, this limitation does not apply
if it is created through non-MHD effects such as the chiral magnetic effect.
One exception is the chiral vortical effect \cite{Tashiro:2012mf},
but since the chiral asymmetry is expected to be set by the temperature,
chiral effects will be constrained as explained in \Eq{argument} of the introduction.
This now seems to be excluded by the observations of blazar spectra,
which are in agreement with the conclusions of Ref.~\cite{Wagstaff:2014fla}.

\begin{acknowledgments}
It is our pleasure to thank Andrey Beresnyak, Alexey Boyarsky,
Ruth Durrer, Arthur Kosowsky, Andrii Neronov, and Oleg Ruchayskiy
for useful discussions.
We thank NORDITA for hospitality and support during the course of this
work. TK also thanks the High Energy and Cosmology division and the
Associate Membership Program at International Center for Theoretical
Physics (ICTP) for hospitality and partial support.
TV also thanks Institute for Advanced Study, Princeton for hospitality
while this work was being completed.
Support through the NSF Astrophysics and Astronomy Grant (AAG) Program
(grants AST1615940 \& AST1615100),
the Research Council of Norway (FRINATEK grant 231444),
the Swiss NSF SCOPES (grant IZ7370-152581), and the Georgian Shota Rustaveli
NSF (grant FR/264/6-350/14) are gratefully acknowledged.
TV is supported by the U.S. Department of Energy Award DE-SC0018330 at ASU.
We acknowledge the allocation of computing resources
provided by the Swedish National Allocations Committee at the Center for
Parallel Computers at the Royal Institute of Technology in Stockholm.
This work utilized the Janus supercomputer, which is supported by the National
Science Foundation (award No.\ CNS-0821794), the University of Colorado
Boulder, the University of Colorado Denver, and the National Center for
Atmospheric Research. The Janus supercomputer is operated by the University of
Colorado Boulder.
\end{acknowledgments}

\appendix
\section{Comparison with the standard MHD equations}
\label{ComparisonStandardMHD}

\begin{figure}[t!]\begin{center}
\includegraphics[width=\columnwidth]{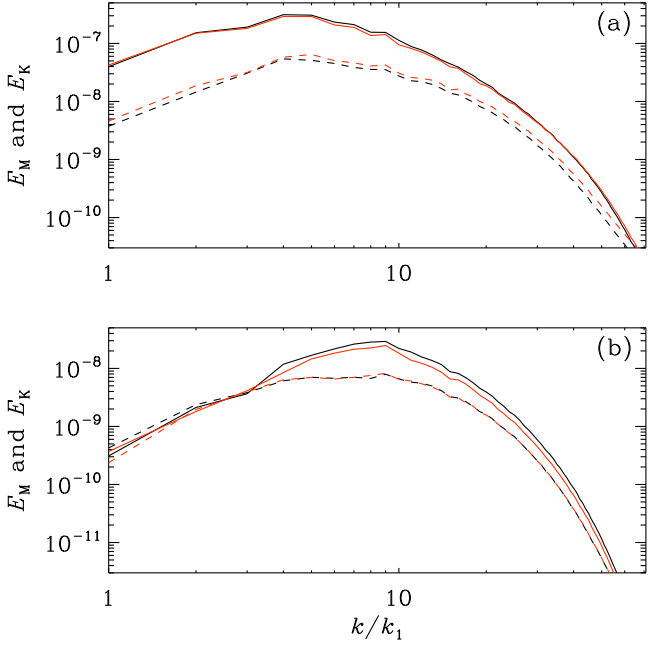}
\end{center}\caption[]{
Magnetic energy spectra (solid lines) and kinetic energy spectra
(dashed lines) for decaying MHD turbulence.
Black (red) lines are for the relativistic (nonrelativistic)
equation of state with (a) $Q_\star=1$ and (b) $Q_\star=0.1$.
}\label{pcomp_spec}\end{figure}

The purpose of this appendix is to contrast \Eqss{dlnrhodt}{mhd3}
with the usual MHD equations for an isothermal gas, i.e.,
\begin{equation}
{\partial\ln\rho\over\partial t}=
-\left(\nab\cdot\uu+\uu\cdot\nab\ln\rho\right),
\end{equation}
\vspace{-6mm}
\begin{eqnarray}
{\DD\uu\over\DD t}\!\!&=&\!\!
-{1\over3}\nab\ln\rho
+{1\over\rho}\JJ\times\BB+{2\over\rho}\nab\cdot\left(\rho\nu\SSSS\right),
\end{eqnarray}
\vspace{-6mm}
\begin{equation}
{\partial\BB\over\partial t}=\nabla\times(\uu\times\BB-\eta\JJ).
\label{mhd3appendix}
\end{equation}
In \Fig{pcomp_spec} we show a comparison of magnetic and kinetic
energy spectra for a low resolution version of Run~I
for the relativistic and nonrelativistic equation of state.
(This run is identical to Run~A of Ref.~\cite{Brandenburg:2017rnt}.)
Note that the magnetic energy spectra are virtually the same,
but the kinetic energy is slightly (factor $4/3$) less in the
relativistic case where $Q_\star=1$; see panel (a).
For the case where $Q_\star=0.1$, the magnetic energy is slightly
(factor $4/3$) larger; see panel (b).

\section{The resulting magnetic field characteristics}
\label{characteristics}

Accounting for the scaling laws obtained for the runs summarized in
\Tab{TSummary}, the (comoving) correlation length and the mean (comoving)
magnetic energy density at time $t$ for $i$th run are given as
\begin{equation}
\xi^{(i)} = \xi^{(i)}_\star \left( {t \over t_\star} \right)^{q_i},
~~~~~
{\cal E}^{(i)}_{\rm M} = {\cal E}^{(i)}_{\rm M\star} \left( {t \over t_\star}\right)^{-p_i}.
\end{equation}
Correspondingly, the magnetic field rms amplitude is
\begin{equation}
B^{(i)}_{\rm rms} = B^{(i)}_{\star, {\rm rms}} \left( {t \over t_\star} \right)^{-p_i/2}.
\end{equation}
Let us consider MHD turbulence decay laws that conserve different
invariants during the turbulent decay process.
In this case the scaling exponents can be calculated using \Tab{TSum},
where $\beta = p/q-1$ can be used as subscript instead of the ``i".
Hence we use $p_\beta$ and $q_\beta$ with $\beta =1,2,4$ for
nonhelical and partially helical fields and $p_0=q_0=2/3$ for the
case of fully helical decay.

If the initial magnetic fields are only partially helical, the first evolutionary
stage consists of the field reaching towards maximal helicity.
During this period, the growth of the correlation length is slower:
$\sim t^{1/2}$ for nonhelical compared to $\sim t^{2/3}$ for fully helical
case in the magnetically dominant scenarios. Also, in this period the mean
magnetic energy density decay is faster: $\sim t^{-1}$ for nonhelical compared to
$\sim t^{-2/3}$ for fully helical case in the magnetically dominant scenarios.
The fractional helicity grows during the turbulence decay process and
reaches the state with maximal helicity in time \cite{Tevzadze:2012kk},
\begin{equation}
t_{\rm hel} = t_\star (\epsilon_{\rm M,\star})^{-1/q_\beta} ~,
\end{equation}
where $\epsMz=\epsM(t_\star)$ and $\epsM$ is defined in Eq.~(\ref{Bikk}).

\begin{figure}[t!]\begin{center}
\includegraphics[width=\columnwidth]{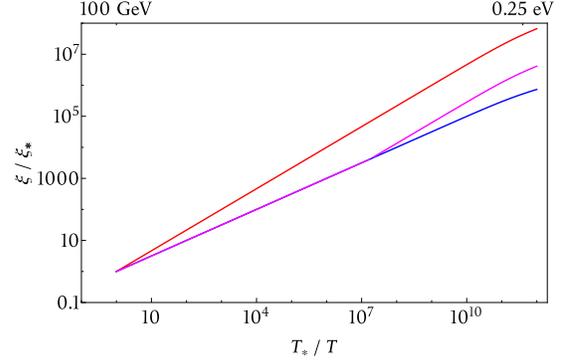}\end{center}
\caption{
Ratio of correlation lengths, $\xiM/\xiMz$, for magnetically dominant
cases for nonhelical (blue), fully helical (red), and fractionally
helical ($\sigma_\star=0.03$; magenta) cases.}
\label{FigAppB1}
\end{figure}

The generation of the magnetic (and/or velocity) field occurs deep in the radiation dominated epoch during which $a\propto t$
(i.e., the conformal time) while the ending
evolution proceeds during the matter dominated epoch when $a\propto t^2$.
To compute the magnetic field characteristic scales at recombination
$t_{\rm rec}$, such as the rms magnetic field amplitude $B_{\rm rec}$ and the
correlation length $\xi_{\rm rec}$, we first calculate the correlation length and the rms
magnetic field when the fully helical state is reached:
\begin{equation}
\xi_{\rm hel} = \xi_{\star} \left( {t_{\rm hel} \over t_{\star}} \right)^{{q_\beta}}, ~~~~~
B_{\rm hel} = B_{\star} \left( {t_{\rm hel} \over t_{\star}}\right)^{-{p_\beta}/2},
\end{equation}
where ${q_\beta}$ ($p_\beta$) is the correlation length scale growth
(the mean magnetic energy density decay) index during the first partially
helical stage: $\beta=1,2,4$.
If the fully helical stage is reached before recombination, the correlation length and the
rms magnetic field at recombination can be calculated as follows:
\begin{equation}
\xi_{\rm rec} = \xi_{\rm hel} \left( {t_{\rm rec} \over t_{\rm hel}} \right)^{{q_0}}, ~~~~~
B_{\rm rec} = B_{\rm hel} \left( {t_{\rm rec} \over t_{\rm hel}} \right)^{-{p_0}/2}
\end{equation}
with $q_0$ ($p_0$) referring to the correlation length (the mean magnetic energy) growth (decay) index during the second helical stage.
It is easy to see that
\begin{eqnarray}
\xi_{\rm rec} &=& \xi_{\star} \left( {t_{\rm rec} \over t_{\star}} \right)^{q_0} \left(\epsilon_{\rm M\star}\right)^{-(q_\beta-q_0)/q_\beta} ~,\nonumber \\
B_{\rm rec} &=& B_{\star} \left( {t_{\rm rec} \over t_{\star}} \right)^{-p_0/2} \left(\epsilon_{\rm M\star}\right)^{(p_\beta-p_0)/2q_\beta}.
\label{rec_scaling}
\end{eqnarray}
Recalling the definition of the $\beta$ parameter and the $pq$ equilibrium condition
(see Sec.~\ref{simparams})
we can express the scaling exponents as follows:
\begin{eqnarray}
q_\beta = {2 \over \beta+3 } ~, \quad
p_\beta = {2 \over \beta+3 } (\beta+1) ~. \label{pq_beta}
\end{eqnarray}
Hence Eqs.~(\ref{rec_scaling}) and (\ref{pq_beta}) show that the
ratio of the correlation length and the mean magnetic field amplitude
at recombination does not depend on the (fractional) helicity of the
initial magnetic field\footnote{Note that in the case of initial kinetic
helicity (which leads to $p\approx q\approx 0.5$; see the square in
the Fig.~\ref{Cons_2}), $B_{\rm rec}/\xi_{\rm rec}$ does depend on the
initial magnetic helicity $\epsM$.} $\epsilon_{\rm M\star}$ or the $\beta$
parameter itself:
\begin{equation}
{B_{\rm rec} \over \xi_{\rm rec}} = {B_{\star} \over \xi_{\star}} \left( {t_{\rm rec} \over t_{\star}} \right)^{-1} ~.
\end{equation}
This helps to set a common recombination limit for different types of
turbulent decay.
On the other hand, the mean magnetic field amplitude and the corresponding
correlation length of the nonhelical or weakly helical fields that do not
have sufficient time to reach a fully helical state before recombination
can be calculated as:
\begin{equation}
B_{\rm rms} = B_\star \left({\xiM \over \xi_\star}\right)^{-(\beta+1)/2} ~.
\end{equation}
\FFig{Cons_2} shows the evolution of the mean turbulent magnetic
field amplitude with respect to the correlation length $\xiM$ for
different classes of MHD turbulence.
Initial values ($\xi_\star$, $B_\star$) correspond to the maximal values
set by BBN constraints at the EWPT epoch.

We provide some numerical estimates of the growth of correlation lengths for the magnetically dominant case. We take
$\xi_{{\rm M},\star}$ to be the maximum comoving Hubble radius at the epoch of electroweak phase transition, given by \eqref{lambda-max}, and note
that at recombination, the temperature was $\sim0.25\,\mathrm{eV}$. The correlation length evolution relations stated above
can be plotted as in \Fig{FigAppB1}.

We have used the fact that the conformal time is expressed in terms
of the scale factor $a_\mathrm{eq}$ at the epoch of matter-radiation
equality and the fractional matter density $\Omega_{m,0}$ as
\begin{equation}
t(a)=\frac{2}{\sqrt{\Omega_{m,0}}H_0}\left[\sqrt{a_\mathrm{eq}+a}-\sqrt{a_\mathrm{eq}}\right],
\end{equation}
and that the effective degrees of freedom of the particle species are roughly constant throughout.

\bibstyle{aps}
\bibliography{paper}

\end{document}